\documentclass[fleqn,usenatbib]{mnras}

\usepackage{newtxtext,newtxmath}
\usepackage[T1]{fontenc}
\usepackage{ulem}
\DeclareRobustCommand{\VAN}[3]{#2}
\let\VANthebibliography\thebibliography
\def\thebibliography{\DeclareRobustCommand{\VAN}[3]{##3}\VANthebibliography}

\usepackage{graphicx}	% Including figure files
\usepackage{amsmath}	% Advanced maths commands
\usepackage{multicol}
\usepackage{subcaption}
\usepackage{comment}
\newcommand{\roughly}{\mathchar"5218\relax\,} % Different from \sim in spacing

\title[BBH Clustering with $k$NN distributions]{Spatial clustering of gravitational wave sources with $k$-nearest neighbour distributions}

\author[K. Gupta and A. Banerjee]{
Kaustubh Rajesh Gupta$^{1}$\thanks{E-mail: kaustubh.gupta@students.iiserpune.ac.in} and Arka Banerjee$^{1}$\thanks{E-mail: arka@iiserpune.ac.in} \\
$^{1}$Department of Physics, Indian Institute of Science Education and Research, Homi Bhabha Road, Pashan, Pune 411008, India\\
}

\date{Accepted XXX. Received YYY; in original form ZZZ}

\pubyear{2024}

\begin{document}
\label{firstpage}
\pagerange{\pageref{firstpage}--\pageref{lastpage}}
\maketitle

\begin{abstract}
    We present a framework to quantify the clustering of gravitational wave (GW) transient sources and measure their spatial cross-correlation with the large-scale structure (LSS) of the universe using the $k$-nearest neighbour ($k$NN) formalism. As a first application, we measure the nearest-neighbour distributions of $53$ suitably selected Binary Black Hole (BBH) mergers detected in the first three observation runs of LIGO-Virgo-KAGRA and cross-correlate these sources with $\roughly 1.7 \times 10^7$ galaxies and quasars from the WISE$\times$SuperCOSMOS all-sky catalogue. To determine the significance of the clustering signal while accounting for observational systematics in the GW data, we create 135 realisations of mock BBHs that are statistically similar to the observed BBHs but spatially unclustered. We find no evidence for spatial clustering or cross-correlation with LSS in the data and conclude that the present sky localisation and number of detections are insufficient to get a statistically significant clustering signal. Looking forward, the statistically large number of detections and the significant improvements in sky localisations expected from future observing runs of LIGO (including LIGO India) and the next generation of GW detectors will enable measurement of the BBH-LSS cross-correlation and open a new window into cosmology.
\end{abstract}


%%%%%%%%%%%%%%%%%%%%%%%%%%%%%%%%%%%%%%%%%%%%%%%%%%

%%%%%%%%%%%%%%%%% BODY OF PAPER %%%%%%%%%%%%%%%%%%

\section{Introduction}
\label{sec:intro}

The clustering of objects that trace structure formation in the universe contains a wealth of information that can be used to test the standard model of cosmology and constrain its parameters \cite[see, e.g.,][]{des_collaboration_dark_2022, des_collaboration_dark_2022-2, des_collaboration_dark_2022-1, des_collaboration_dark_2023, dvornik_kids-1000_2023, amon_consistent_2023, miyatake_hyper_2023, fumagalli_a_cosmological_2024}. The detection of gravitational waves by the LIGO-Virgo-KAGRA (LVK) collaboration \citep{ligo_scientific_collaboration_and_virgo_collaboration_observation_2016} has unveiled potential new tracers of structure formation in the form of merging binaries of compact stellar remnants such as black holes and neutron stars \citep{ligo_scientific_collaboration_gwtc-3_2023}.

Gravitational waves allow a direct measurement of the luminosity distance to their sources \cite[see, e.g.,][]{schutz_determining_1986, Holz_2005, 10.1063/PT.3.4090} without the need for a hierarchical distance ladder or an empirical calibration process, with the only fundamental assumption being that general relativity is valid, making merging compact binaries `standard sirens'. Hence, gravitational waves provide a mechanism to study the expansion history of our universe if the source redshifts can be estimated. This led \cite{schutz_determining_1986} to suggest that merging compact binaries can be used to constrain the Hubble-Lema\^{\i}tre parameter $H_0$, which characterises the present-day rate of expansion of the universe.

Since the redshift of a merging binary cannot be inferred directly from its gravitational waves, many techniques have been developed in the literature to measure $H_0$ using additional astrophysical observations (see \cite{https://doi.org/10.1002/andp.202200180} for a recent review). For example, the `bright siren' method uses direct electromagnetic counterparts of the merger events to obtain redshifts \citep{abbott_gravitational-wave_2017}. In contrast, the statistical dark siren method (see \cite{Gair_2023} for a review) uses galaxy surveys to identify potential hosts of the merger events inside the localisation volumes provided by gravitational wave observations \citep{Soares-Santos_2019, palmese_statistical_2020, abbott_gravitational-wave_2021, alfradique_dark_2024}. \cite{PhysRevD.77.043512} proposed a method that uses galaxy clustering to extract redshift information for a sample of merger events in a statistical sense to estimate $H_0$ without needing to identify host galaxies for individual merger events. Methods have also been proposed that try to estimate the redshift of gravitational wave sources by breaking the mass-redshift degeneracy in gravitational wave analyses; this can be achieved, for example, by constraining the neutron star tidal deformability or by combining features in the mass distribution and redshift evolution of merger rate (the so-called `spectral siren method'), to measure the source masses \citep{Farr_2019, farr_future_2019, PhysRevD.104.062009, PhysRevLett.129.061102, mancarella2022cosmology, abbott_constraints_2023}. 

All of these methods, however, have various drawbacks \citep{https://doi.org/10.1002/andp.202200180}. The bright siren method relies on detecting rare events accompanied by electromagnetic counterparts and possibly only applies to binary neutron star (BNS) mergers. The dark siren method suffers from difficulties due to large localisation volumes of gravitational wave events and is susceptible to potential biases in the inference of $H_0$ due to the incompleteness of galaxy catalogues \citep{trott2022challenges}. The mass-redshift degeneracy method is model-dependent; uncertainties in the modelling of the neutron star equation of state, or a wrong model of the binary merger rate, can introduce systematic biases in the cosmological inference (see section 2.3.3 of \cite{https://doi.org/10.1002/andp.202200180} and references therein). 

If star-forming regions follow the fluctuations in the underlying cosmological matter field, merging compact binaries are expected to be inherently clustered and spatially correlated with other tracers such as galaxies and galaxy clusters \citep{Scelfo_2018}. The strength of the cross-correlation between sources of gravitational waves and the large-scale structure of the universe is sensitive to cosmological parameters. It can, therefore, be used as an independent probe of the Hubble-Lema\^{\i}tre constant \citep{PhysRevD.93.083511, Bera_2020, PhysRevD.103.043520, mukherjee_cross-correlating_2022}, after marginalizing over the other relevant parameters. As long as the gravitational wave sources and the galaxy sample trace fluctuations in the same underlying density field, a measurement of their spatial cross-correlation does not require uniquely identifying the source of each merger event (\cite{Fang_2020} have a similar discussion in the context of cross-correlations between high-energy neutrinos and large-scale structure). Therefore, the cross-correlation method of determining $H_0$ does not suffer from biases due to the incompleteness of the galaxy catalogues used \cite[see][for a systematic study]{Bera_2020}. Moreover, this method does not require any direct assumptions about the population properties of the merging binaries. Therefore, it measures $H_0$ nearly independently of merging binary population models\footnote{It is to be noted, however, that the clustering of gravitational wave sources is also expected to be a function of bias parameters that model the tracer-matter connection. One needs to marginalise over these parameters to obtain constraints on cosmological parameters \cite[see, e.g.,][and references therein]{10.1093/mnras/stac193, 2023arXiv230518003P}. Since the bias parameters are controlled by the source population properties \citep{PhysRevD.94.023516, Adhikari_2020, 2023arXiv230518003P, 2023arXiv231203316V}, an indirect dependence on population models is introduced in the measurement of $H_0$.}.

Since the various techniques of measuring $H_0$ using gravitational wave sources discussed above do not utilise information in the temperature fluctuations of the cosmic microwave background (CMB) \citep{ade_planck_2016, aghanim_planck_2020} or cosmic distance ladder distance measurements from supernovae and other standard candles \citep{Riess_2018, riess_large_2019, riess_expansion_2020, wong_h0licow_2020, riess2022comprehensive}, measuring the cross-correlation between gravitational wave sources and the large-scale structure of the universe is important in the context of the so-called Hubble Tension (see \cite{hu_hubble_2023} and \cite{valentino_realm_2021} for comprehensive reviews). In addition to cosmology, these cross-correlation measurements can also be used to study the astrophysical origins and formation channels of gravitational wave sources \cite[see, e.g.,][]{PhysRevD.94.023516, Scelfo_2018, scelfo_exploring_2020, Adhikari_2020, 2023arXiv231216289G}. With the third generation of gravitational wave detectors likely to bring in $\roughly 10^5$ more detections per year \citep{iacovelli_forecasting_2022, borhanian_listening_2022}, measuring the clustering of these objects and modelling it as a function of cosmological parameters will, therefore, play an essential role for both precision cosmology and compact binary astrophysics in the coming decade.

There have been a few attempts to measure the angular two-point correlation function \citep{cavaglia_two-dimensional_2020, zheng_angular_2023} and the angular power spectrum \citep{zheng_angular_2023} of the currently detected LVK events; to determine the level of anisotropy in their sky distribution \citep{PhysRevD.102.102004, PhysRevD.107.043016}; and to measure their spatial cross-correlation with galaxy catalogues \citep{mukherjee_cross-correlating_2022}. However, a statistically significant detection of anisotropy or clustering has not yet been achieved. Studies using forecasts for future detectors have also been performed \citep{PhysRevLett.116.121302, Scelfo_2018, calore_cross-correlating_2020, Libanore_2021, Libanore_2022, 2023JCAP...06..050B, vijaykumar_probing_2023, 2023arXiv231216289G}, primarily focusing on two-point summary statistics. 

In this paper, we present a framework to quantify the clustering of gravitational wave sources and their spatial cross-correlation with large-scale structure catalogues using the $k$-nearest-neighbour distributions \citep{banerjee_nearest_2021} as summary statistics. The nearest-neighbour measurements are sensitive to all $N$-point correlation functions of the tracers and hence are a much more powerful probe of clustering, compared to the two-point function, on scales where the underlying matter field is not well-approximated as a Gaussian random field, and the effect of gravitational nonlinearities cannot be neglected \citep{banerjee_nearest_2021, banerjee_cosmological_2021, banerjee_tracer-field_2023}. Application of these statistics could, in principle, lead to a detection of the clustering signal from the same datasets used in previous two-point analyses. To enable this new analysis, we extend the $k$NN formalism, originally presented for 3D clustering in cartesian coordinates, to angular clustering in the sky. As a first application to data, we compute the auto-correlation of a suitable subset of the binary black holes (BBHs) detected in the first three observing runs of LVK and their cross-correlation with the WISE$\times$SuperCOSMOS all-sky survey. We also compare the results of the two-point and nearest-neighbour analyses. Although we focus on BBHs in this work, our framework can easily be extended to study the clustering of other gravitational wave transients like binary neutron stars and neutron star-black hole binaries.

The rest of the paper is structured as follows. We describe the data used in this study in section~\ref{sec:data}. In section~\ref{sec:math}, we develop the mathematical formalism for clustering statistics and discuss how to compute them numerically. At the end of section~\ref{sec:math}, we present an illustrative example that demonstrates the potential boost in the clustering signal of sparsely sampled tracers expected from the nearest-neighbour measurements on small spatial scales over the two-point summary statistics. Having motivated the $k$NN formalism, in section~\ref{sec:methods}, we describe our methods: the null hypothesis, our procedure to test the null hypothesis, how we determine the statistical significance of the clustering signal, and our strategy to deal with observational selection biases in the data. We present our results in section~\ref{sec:results} and discuss some interesting aspects of this study. Finally, we summarise, draw conclusions, and discuss possible future directions in section~\ref{sec:conc}. Some additional material is presented in the appendix. 

\section{Data}
\label{sec:data}

In this section, we discuss the data used in this study. We describe the gravitational wave events selected for this work in section~\ref{subsec:GWdata}. Typically, a mock catalogue of unclustered data points (known as `randoms' in the literature) is required to get a reliable measurement of the statistical significance of the clustering signal in the presence of observational selection biases \cite[see., e.g.,][]{10.1093/mnras/stac1551}. In section~\ref{subsec:mockdata}, we motivate this requirement for the specific case of gravitational wave data, discuss the procedure to create the unclustered catalogue, and present the resulting mock data. Finally, we describe the large-scale structure catalogue used for cross-correlating the BBHs in section~\ref{subsec:Galdata}.

\subsection{Gravitational Wave Events}
\label{subsec:GWdata}

This work uses the compact binary merger events detected in the first three observing runs of LIGO-Virgo-KAGRA, as reported in \cite{ligo_scientific_collaboration_gwtc-3_2023}. Following \cite{zheng_angular_2023}, from this parent set of $\roughly 80$ events, we select the events detected with a false alarm rate (FAR) less than 1 per year and crossed a detection threshold of network matched-filtered signal-to-noise ratio (SNR) greater than 10. Since we are interested in binary black holes (BBHs), we further restrict our sample to those events that have a probability of being a BBH merger greater than 0.5\footnote{The classification probabilities were calculated using the \texttt{GW} package of \texttt{PESummary} \citep{hoy_pesummary_2021}.}. 

\cite{zheng_angular_2023} restrict their sample to events detected in all three detectors that were in science mode during the LVK observing period, namely LIGO Livingston, LIGO Hanford \citep{Aasi_2015} and Virgo \citep{Accadia_2012}, to get better-localised events. However, this step removes a significant fraction of the BBHs selected above. Since it is not clear a priori whether the resulting gain in sky localisation accuracy would compensate for the reduction in the BBH sample size, we keep two-detector events in our final sample. To ensure better homogeneity in the sky localisations of the BBHs, we remove all events from our sample detected before the Virgo detector joined the observing run. Note that a non-detection in one of the detectors does carry some information about the location of the merger event in the sky. Hence, the two-detector events not detected in Virgo when it is in science mode are still expected to be better localised than those observed in the absence of Virgo. 
% \arka{Can we point to a reference which motivates these specific choices?} \kaustubh{Has this been answered?}

\begin{figure}
    \centering
    \includegraphics[width=\columnwidth]{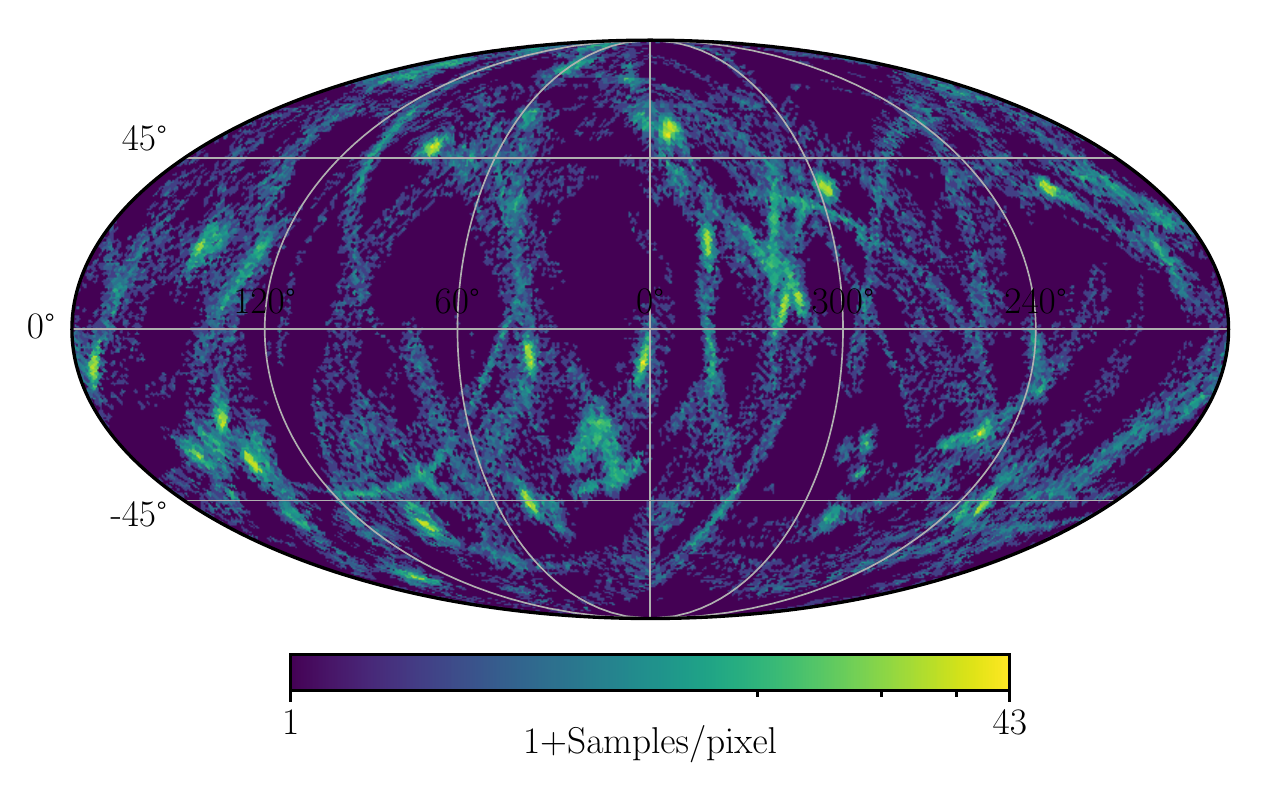}
    \caption{Mollweide projection of the combined skymap of the 53 observed events that constitute our BBH catalogue, in equatorial (J2000) coordinates. Each banana-shaped cloud represents a single BBH. Skymaps were generated through the \texttt{Healpy} package using the parameter estimation posterior samples provided by the LVK collaboration for each event. The colour represents the number of posterior samples per pixel in a logarithmic scale. The HEALPix \texttt{NSIDE} for this map is 64.}
    \label{fig:1}
\end{figure}

Finally, we are left with 53 BBHs that constitute our observed catalogue. Using the parameter estimation posterior samples on declination and right ascension made publicly available by the LVK collaboration, we generate skymaps for each event representing the uncertainty in their localisation in the sky. Figure~\ref{fig:1} shows a combined skymap of all events generated by stacking the individual skymaps. We summarise the properties of the observed BBHs that are most relevant for clustering, namely the distribution of sky localisation areas and observed luminosity distances, in Figure~\ref{fig:2}. Distributions of other BBH properties are summarised in appendix~\ref{sec:A}.

\begin{figure*}
    \centering
    \begin{multicols}{2}
        \subcaptionbox{68\% credible area\label{fig:2a}}{\includegraphics[width=\linewidth]{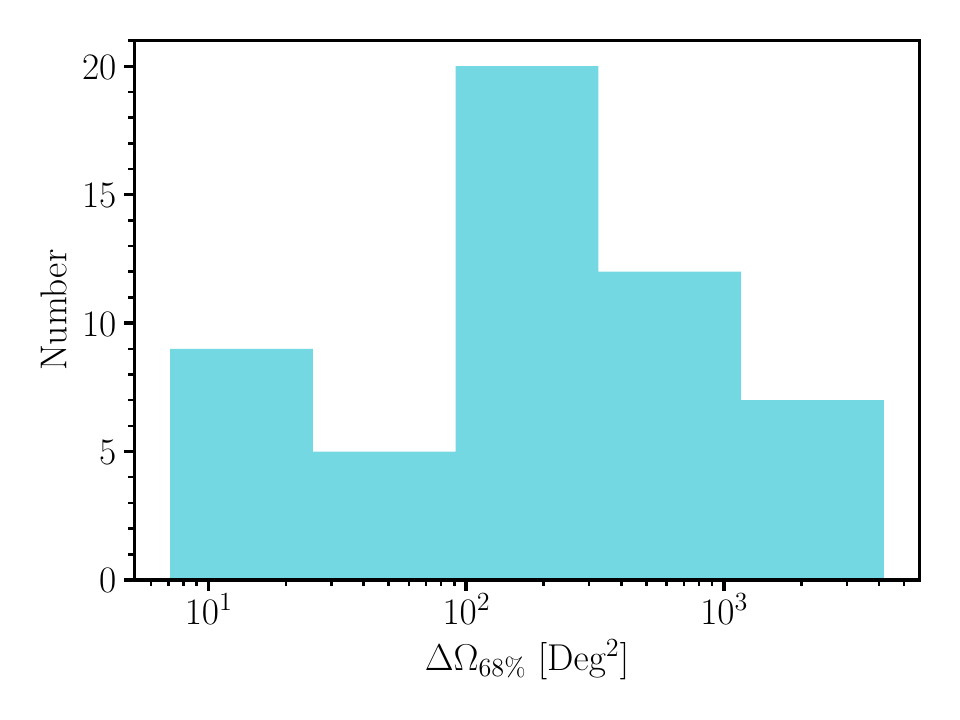}}\par 
        \subcaptionbox{Luminosity distance\label{fig:2b}}{\includegraphics[width=\linewidth]{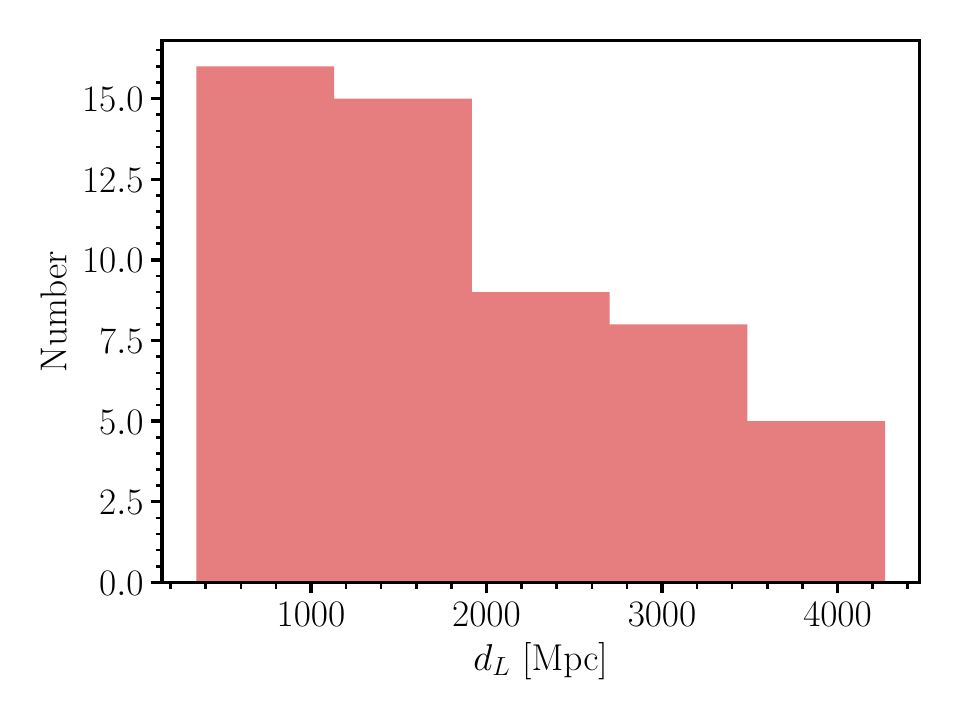}}\par 
    \end{multicols}
    \caption{Distribution of the $1\sigma$ sky localisation uncertainty areas (\textit{left}) and luminosity distances (\textit{right}) of the observed BBH catalogue. The credible areas are computed using the rapid Bayesian localisation code \texttt{BAYESTAR} \citep{singer_rapid_2016}.}
    \label{fig:2}
\end{figure*}

\subsection{Mock BBH Catalog}
\label{subsec:mockdata}

In this section, we describe the procedure for creating the mock BBH catalogue that will serve as the set of `randoms' used for the clustering analysis. At first, it appears that simply distributing points uniformly in the sky should be sufficient, as the resulting data set would be unclustered and uncorrelated with large-scale structure. This would be a valid approach if we had perfect observations of a sample of BBHs representative of the entire BBH population in the universe. Unfortunately, due to the limited sensitivity of the current gravitational wave detectors, the data are plagued with selection biases and systematic effects; the BBHs selected for this study do not constitute a representative sample of the population\footnote{It should be noted, however, that the third generation of gravitational wave detectors is expected to detect almost all BBH mergers in the universe up to a very high redshift \citep{Hall_2019, iacovelli_forecasting_2022, borhanian_listening_2022}.}. Furthermore, the detectors are not equally sensitive to all regions in the sky; each detector is most sensitive to the merger events that go off directly on top of it, i.e., perpendicular to the plane of detector arms. This is a consequence of the transverse nature of gravitational waves. As a result, there is a selection function in the sky for the detector network as a whole (see \cite{Chen_2017} for example). These observational systematics have to be carefully folded into the clustering analysis to avoid getting biased or spurious signals. 

One way of mitigating the selection biases outlined above is to create realistic mock BBHs that reproduce the properties of the observed BBH sample in a statistical sense but which are inherently unclustered and spatially uncorrelated with the large-scale structure of the universe. Such a mock data set allows us to naturally incorporate the effects of observational biases on the clustering measurements.

We follow a procedure outlined in \cite{zheng_angular_2023} to create our mock catalogue. First, we distribute a population of BBH merger events isotropically in the sky by sampling their locations from a uniform distribution ($\mathcal{U}$), which translates to drawing their right ascension ($\alpha$) from $\mathcal{U}(0, 2\pi)$ and sine of declination ($\sin\delta$) from $\mathcal{U}(-1, 1)$. We next draw their source parameters from the population distributions inferred by the LVK collaboration \citep{ligo_scientific_collaboration_population_2023} as implemented by the \texttt{GWPopulation} package\footnote{\url{https://colmtalbot.github.io/gwpopulation/}}. These are as follows:
\begin{enumerate}
    \item \texttt{Power Law + Peak} model for the mass of the primary (heavier) BBH and a power law distribution for the ratio of component masses \citep{talbot_measuring_2018}
    \item power law distribution for redshift evolution of merger rate per unit comoving volume per unit source-frame time \citep{fishbach_does_2018}
\end{enumerate}
The mathematical details of these models are discussed in appendix~\ref{sec:B}. We assume uniform distributions for inclination angle $\iota$ w.r.t. the plane of orbital angular momentum, polarisation angle $\psi$ and phase at coalescence $\Phi_c$ over their allowed physical ranges, i.e., $\iota, \psi \in \mathcal{U}(0, \pi)$ and $\Phi_c \in \mathcal{U}(0, 2\pi)$, and uniformly sample the BBH merger time during the LIGO observation period after Virgo started taking data. For simplicity, we set the black hole spins identically to zero since we do not expect them to affect the clustering properties or the sky localisation uncertainties, which are most relevant to us. We have further checked that including the spins does not affect our analysis; the final mock BBH catalogues with and without spins turned on are statistically similar in all aspects.

After creating the mock BBH population, we determine which events can be `detected' by the current gravitational wave detectors to reproduce the selection biases in the data. Here, we consider a detector network consisting of LIGO Livingston, LIGO Hanford and Virgo, the same network that collected the data for our observational sample. We conduct the following gravitational wave data analysis using the rapid Bayesian localisation code for gravitational wave events, $\texttt{BAYESTAR}$\footnote{We follow a similar procedure to the one outlined in \url{https://lscsoft.docs.ligo.org/ligo.skymap/quickstart/bayestar-injections.html}.} \citep{singer_rapid_2016}. First, we simulate the gravitational wave signals for each BBH using the \texttt{IMRPhenomXPHM} model \citep{pratten_computationally_2021}, which is the same waveform used in the LVK analysis of the data. Next, we inject the simulated signals in stationary Gaussian noise created using analytic estimates for the third observing run power spectral densities for the LIGO Livingston, LIGO Hanford and Virgo detectors, as provided by the \texttt{PyCBC} package\footnote{\url{http://pycbc.org/pycbc/latest/html/index.html}}. Finally, we compute the (phase-maximised) network matched-filtered signal-to-noise (SNR henceforth) for each event and classify the events with an SNR $\geq 10$ as `detections'\footnote{Technically, the false alarm rate (FAR) is a better measure of whether an event should be considered detectable, but computing FARs is computationally expensive, as it requires doing parameter estimation for the full set of injected events. An SNR cutoff of 10 is a reasonable proxy (see \citep{ligo_scientific_collaboration_gwtc-3_2023} or \citep{essick_semianalytic_2023} for more details).}. Once we have the selected events, we use \texttt{BAYESTAR} to localise them. \texttt{BAYESTAR} also returns the estimated luminosity distances and credible intervals for the area of sky localisation uncertainty. 

\begin{figure*}
    \centering
    \begin{multicols}{2}
        \subcaptionbox{Sample realisation 1\label{fig:6a}}{\includegraphics[width=\linewidth]{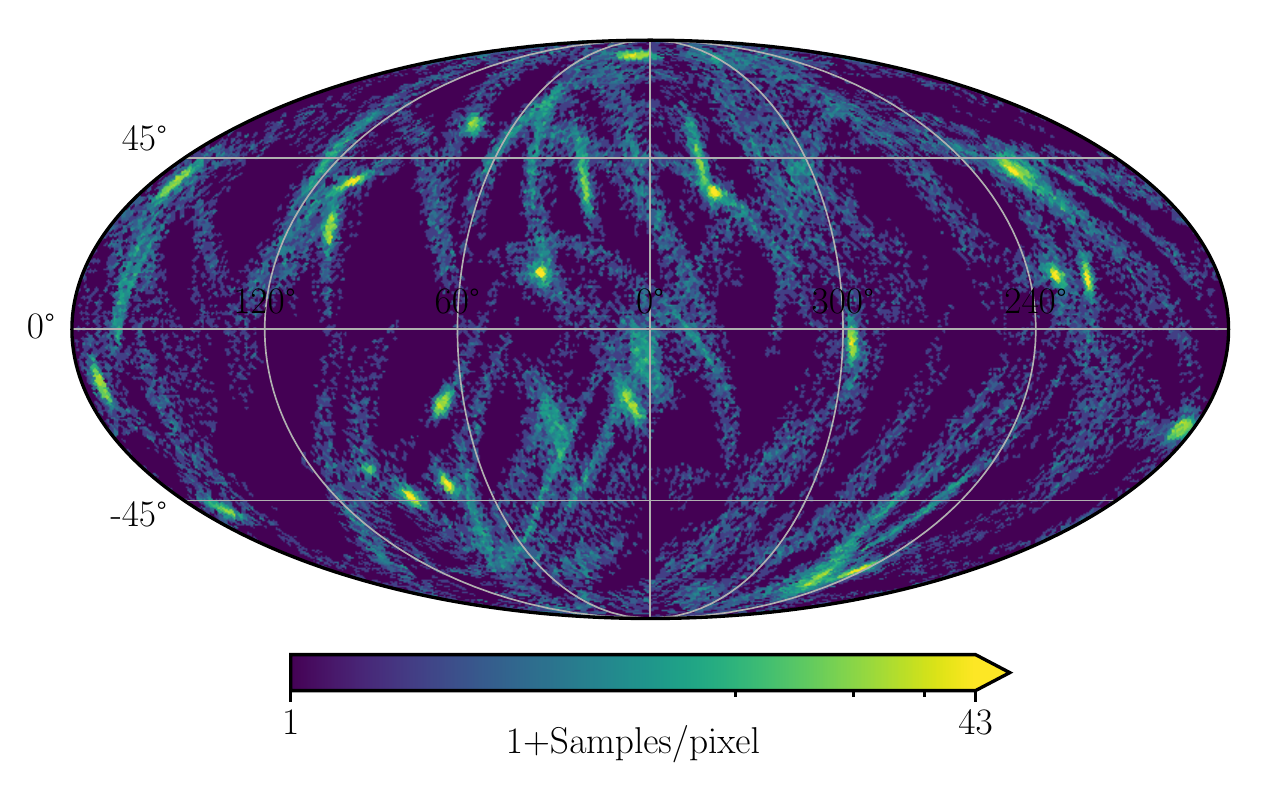}}\par 
        \subcaptionbox{Sample realisation 2\label{fig:6b}}{\includegraphics[width=\linewidth]{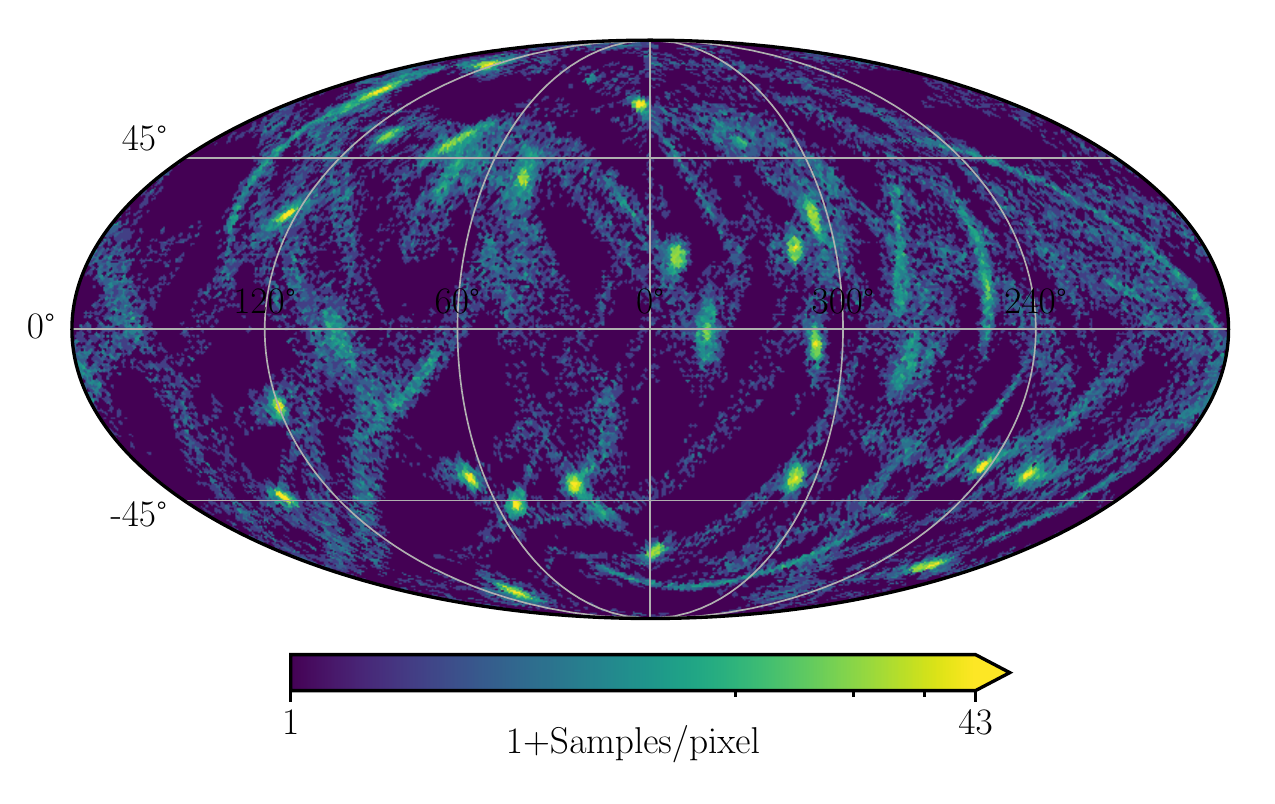}}\par 
    \end{multicols}
    \begin{multicols}{2}
        \subcaptionbox{Sample realisation 3\label{fig:6c}}{\includegraphics[width=\linewidth]{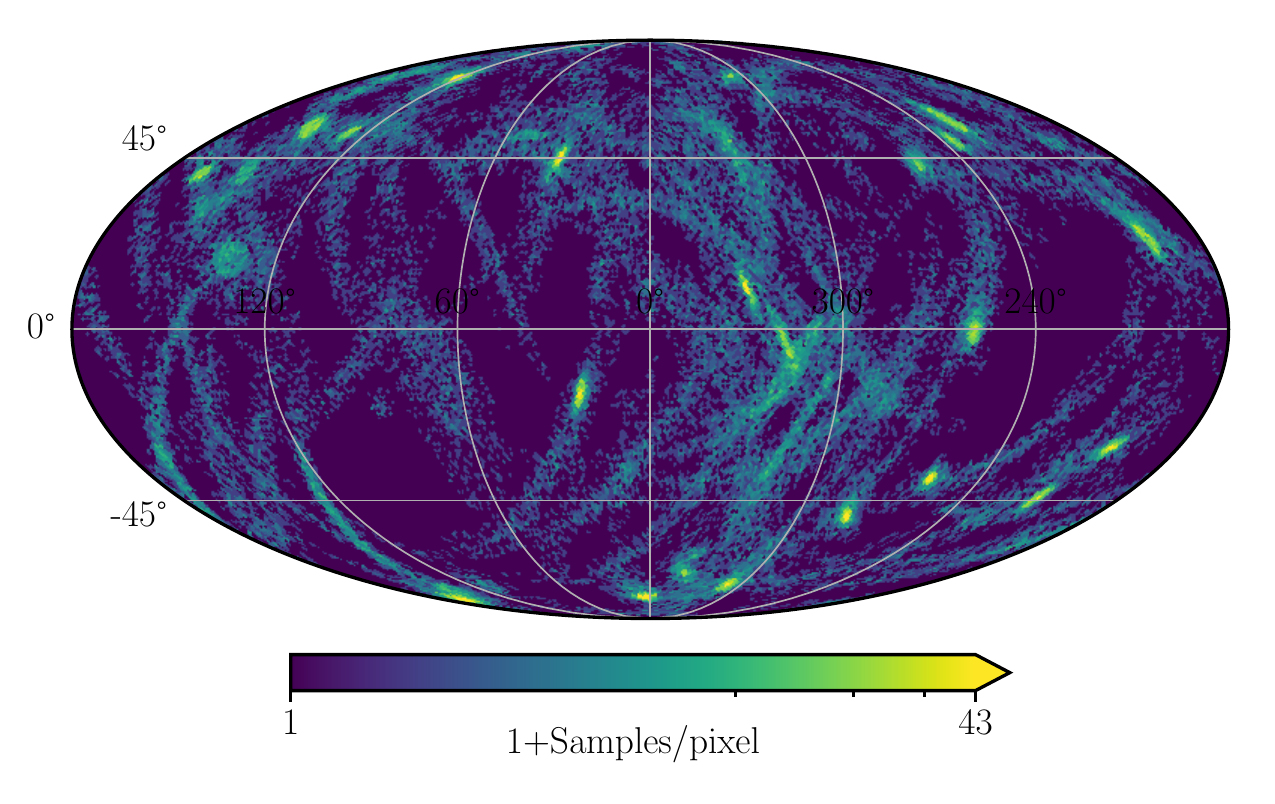}}\par 
        \subcaptionbox{Sample realisation 4\label{fig:6d}}{\includegraphics[width=\linewidth]{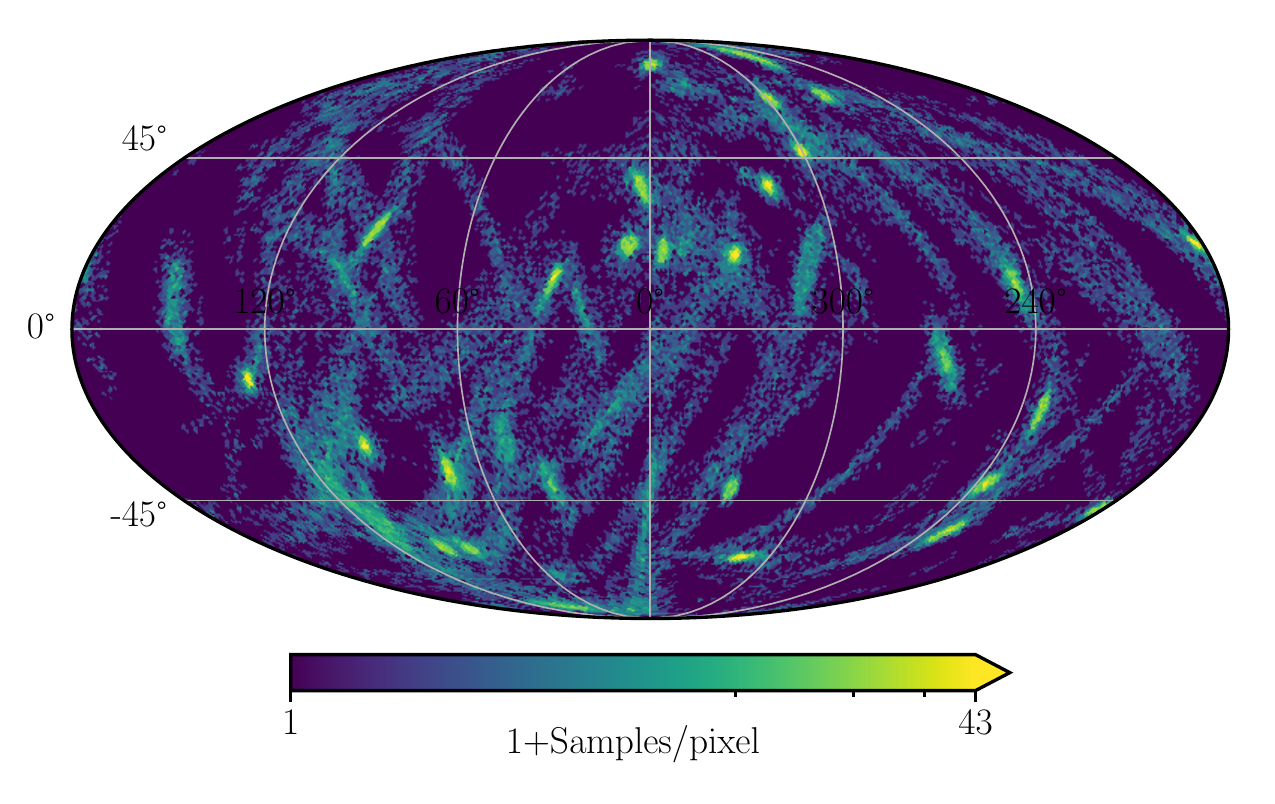}}\par
    \end{multicols}
    \caption{Mollweide projection of the combined skymap of the 4 sample realisations of the mock BBH catalogue in equatorial (J2000) coordinates. Each banana-shaped cloud represents a single BBH. These are visually similar to figure~\ref{fig:1}. Skymaps were generated using \texttt{BAYESTAR}. As before, the colour represents the number of posterior samples per pixel in a logarithmic scale. The HEALPix \texttt{NSIDE} for these maps is 64.}
    \label{fig:6}
\end{figure*}

Using the procedure outline above, we generate a mock catalogue of $135$ realisations of $53$ BBHs each. The combined skymaps of $4$ sample realisations are shown in figure~\ref{fig:6}, with the colour palate and resolution identical to figure~\ref{fig:1} for ease of comparison. The skymaps of the observed and mock BBHs are visually similar. To investigate if the mock catalogue is statistically similar to the observational data, we compare their sky localisation uncertainty and luminosity distance distributions, averaged over the $135$ realisations, with the corresponding distributions measured in the data. The results are shown in figure~\ref{fig:7}. Similar plots for other source properties are presented in appendix~\ref{sec:C}. 

Within the limit of sample variance across the realisations, the mock catalogue is reasonably statistically similar to the observed BBH sample. It does includes more events with higher uncertainty in the sky localisation, as seen from the $\roughly 3\sigma$ deviation from the observed distribution in the second last histogram bin of figure~\ref{fig:7a}. However, this is not expected to bias the clustering results since it would only lead to slightly larger measurement errors for the clustering statistics of each mock realisation. As we discuss in section~\ref{subsec:significance}, the relevant quantity for measuring the significance of the clustering signal is the variance across the realisations, which is independent of the measurement errors on the individual realisations.

\begin{figure*}
    \centering
    \begin{multicols}{2}
        \subcaptionbox{68\% credible area\label{fig:7a}}{\includegraphics[width=\linewidth]{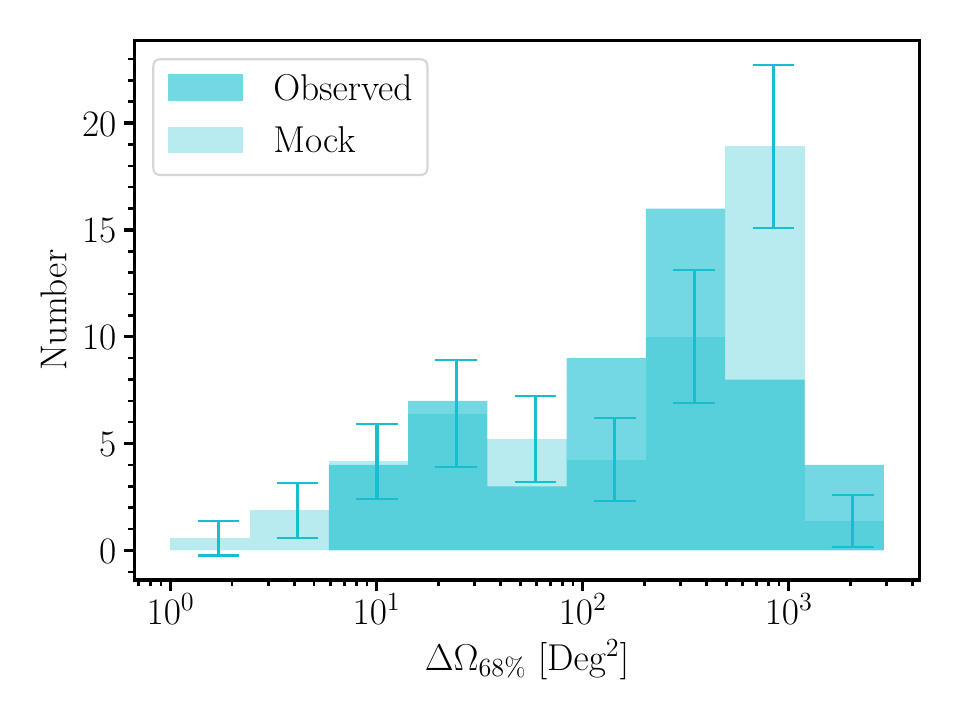}}\par 
        \subcaptionbox{Luminosity distance\label{fig:7b}}{\includegraphics[width=\linewidth]{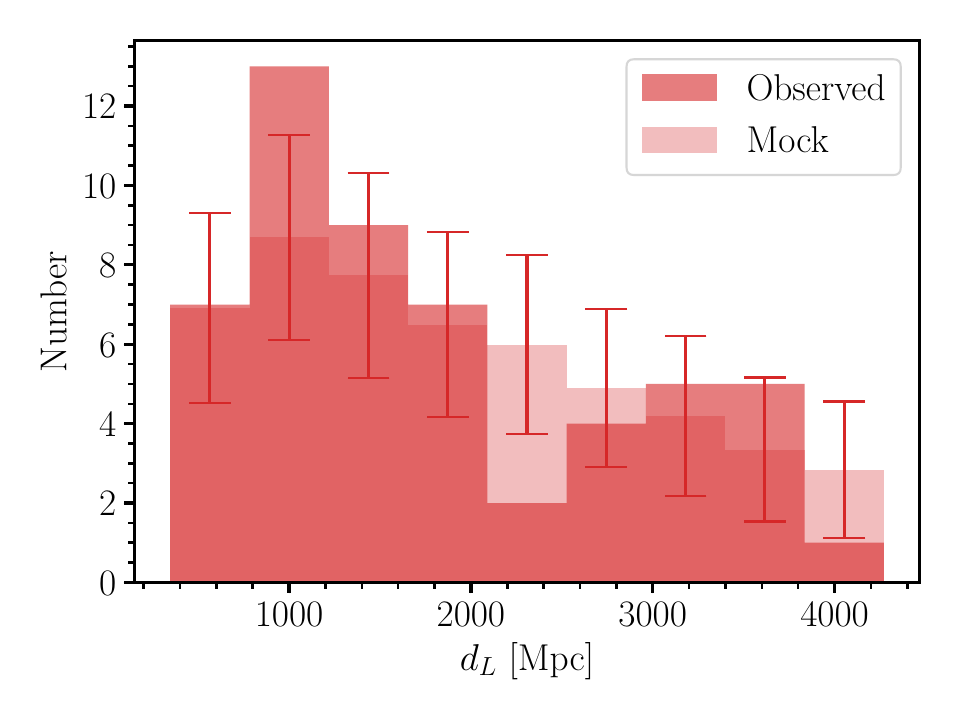}}\par 
    \end{multicols}
    \caption{Distribution of the $1\sigma$ sky localisation uncertainty areas (\textit{left}) and luminosity distances (\textit{right}) of the observed (bold histograms) and mock (light histograms) BBHs. The error bars on the mock histograms show the variance over 135 realisations of the mock catalogue. Except for one or two bins, the mock and the observed data agree reasonably within the error bars, signifying that the mock dataset statistically reproduces the observations.}
    \label{fig:7}
\end{figure*}

\subsection{Galaxy Catalogue}
\label{subsec:Galdata}

We use galaxies and quasars from the publicly available WISE$\times$SuperCOSMOS (hereafter WSC) catalogue \citep{Bilicki_2016}, which is a cross-match between two parent full-sky catalogues: the AllWISE release \citep{cutri_explanatory_2013} from the Wide-field Infrared Survey Explorer (WISE) \citep{wright_wide-field_2010}, a mid-infrared ($\lambda \roughly \mu m$) space survey; and the SuperCOSMOS Sky Survey \citep{hambly_supercosmos_2001}, consisting of data from digitised optical photographic plates taken by the United Kingdom Schmidt Telescope (UKST) in the southern hemisphere\footnote{\url{https://www.roe.ac.uk/ifa/wfau/ukstu/telescope.html}} and the Palomar Observatory Sky Survey-II (POSS-II), in the northern hemisphere \citep{reid_second_1991}. WISE, a NASA space-based mission, surveyed the entire sky in four bands, $W_1 = 3.4 \mu m$, $W_2 = 4.6 \mu m$, $W_3 = 12 \mu m$, and $W_4 = 23 \mu m$, while SuperCOSMOS has data in three optical bands, $B$, $R$, and $I$. The interested reader is referred to \cite{Bilicki_2016} for more details on the WISE and SuperCOSMOS surveys and the cross-matching procedure. We use photometric redshifts for the WSC catalogue provided by \cite{Bilicki_2016}, which have been estimated using the artificial neural network code \texttt{ANNz} \citep{collister_annz_2004}. 

We work with the `SVM' release of the WSC catalogue \citep{krakowski_machine-learning_2016}, which classifies sources into galaxies, stars and quasars using a support vector machines (SVM) learning algorithm. The reason for choosing the SVM catalogue is as follows: in creating the original WSC catalogue, colour cuts were placed that already removed the quasars. Since we expect quasars to trace the large-scale fluctuations in the universe alongside galaxies, removing quasars is unnecessary for a cross-correlation study like ours. We remove the sources classified as `stars' in the SVM catalogue and select the remaining objects to form our raw catalogue. Finally, we remove sources lying in problematic regions in the sky with unreliable data using the publicly available WSC mask to create the final catalogue cross-correlated with the BBH catalogue created in section~\ref{subsec:GWdata}. This process removes regions such as those obscured by the plane of the Milky Way and by the Large and Small Magellanic Clouds (SMC and LMC) and the areas with high stellar contamination (see \cite{Bilicki_2016} for a detailed description of the masking procedure). Some authors \cite[for example,][]{mukherjee_cross-correlating_2022} impose additional colour cuts on $E(B - V)$ and/or on $W_1-W_2$ to mitigate dust extinction and further reduce stellar contamination. However, this increases the purity of the galaxy sample at the cost of completeness, which is undesirable for cross-correlation studies. Since we do not expect nearby stars to correlate with the extragalactic BBHs, we do not impose any additional colour cuts. 

\begin{figure}
    \centering
    \includegraphics[width=\columnwidth]{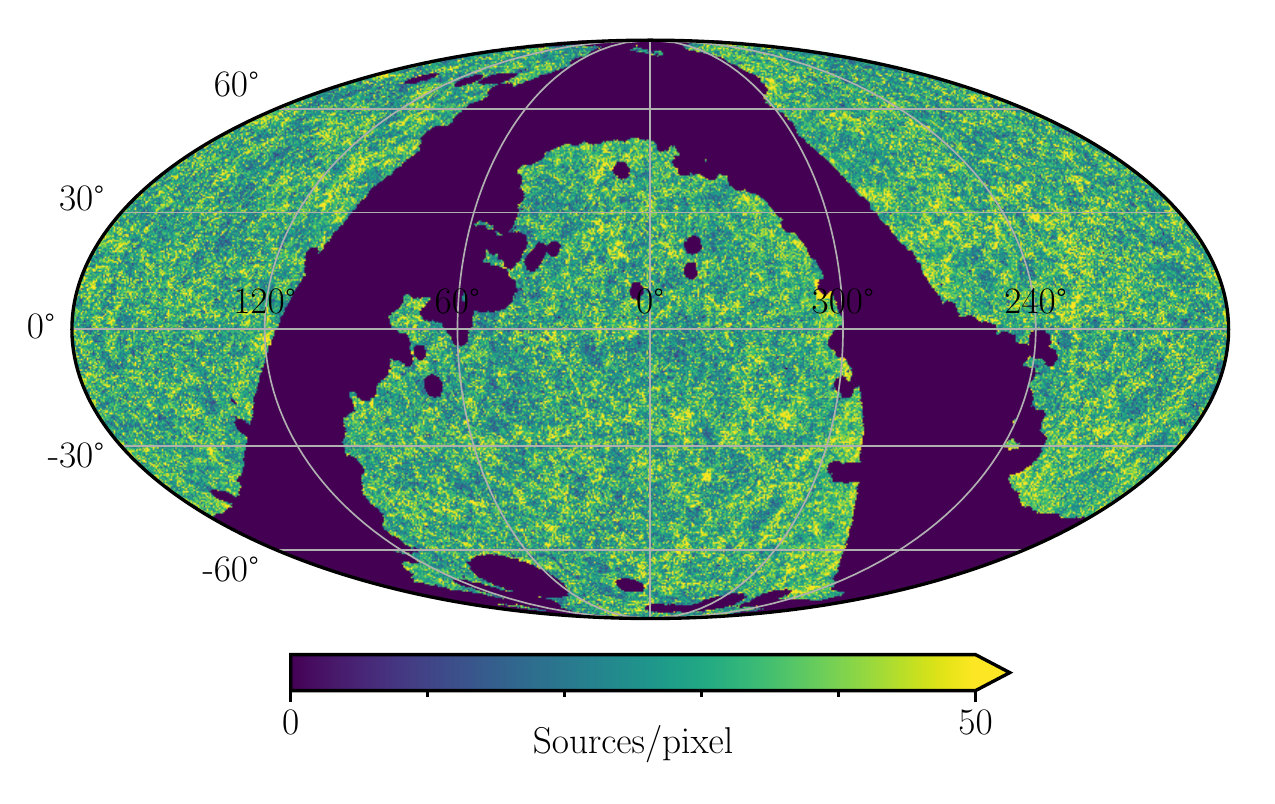}
    \caption{Mollweide projection of the skymap of $\roughly 1.7 \times 10^7$ galaxies and quasars in the WSC catalogue, in equatorial (J2000) coordinates. Skymaps were generated from the sky locations of the sources using the \texttt{Healpy} package. The colour represents the number of sources per pixel on a linear scale. The colour bar has been limited to a maximum of 50 samples per pixel to enhance contrast, which makes it easier to visualise the large-scale structure in the distribution of the galaxies and quasars. The empty navy regions represent regions in the sky with unreliable data and have been masked out. The HEALPix \texttt{NSIDE} for this map is 256.}
    \label{fig:3}
\end{figure}

After masking out regions in the sky with unreliable data, we are left with a catalogue that covers $\roughly 3\pi$ steradians in the sky, corresponding to a sky coverage of $\roughly 68\%$. This makes this catalogue suitable for a cross-correlation study with BBH catalogues, which are inherently all sky since gravitational waves are not susceptible to medium propagation effects\footnote{Note, however, that gravitation waves, like light, are indeed susceptible to weak gravitational lensing \citep{PhysRevD.93.083511, 10.1093/mnras/stz3509}.}. The sky distribution of the WSC galaxies and quasars is shown in figure~\ref{fig:3}.

\begin{figure}
    \centering
    \includegraphics[width=\columnwidth]{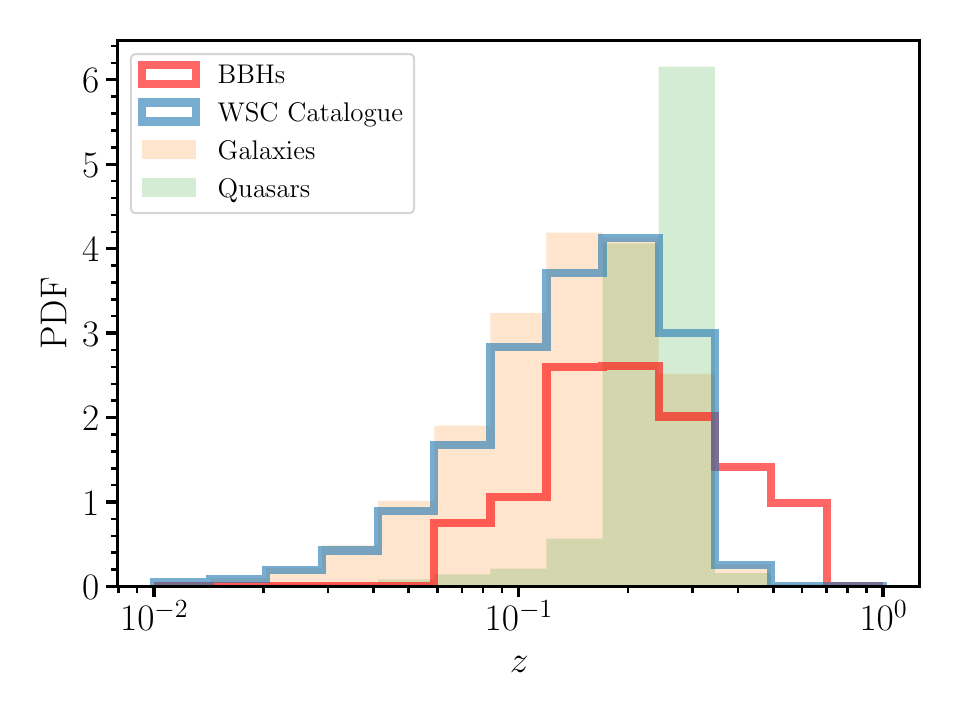}
    \caption{A comparison of the redshift distributions of the observed BBHs (red histogram) and the WSC catalogue sources (blue histogram). There is a significant overlap between the redshifts of the two datasets, which is crucial for conducting cross-correlation studies since we do not expect cosmological fluctuations at different redshifts to be correlated. Also plotted are filled histograms showing the distribution of galaxy (orange) and quasar (green) redshifts separately. The quasars are at a higher redshift on average.}
    \label{fig:4}
\end{figure}

Furthermore, the redshift distribution of the WSC sources significantly overlaps with that of the BBHs selected for this study, as shown in figure~\ref{fig:4}. This is important for cross-correlation studies since we do not expect cosmological fluctuations at different redshifts to be correlated. Note that the redshifts of the BBHs are not direct observables but are computed from the luminosity distances, assuming a cosmological model for the expansion history of the universe. For this dataset, the LVK collaboration \citep{ligo_scientific_collaboration_gwtc-3_2023} assumed a cosmological model consistent with the Planck 2015 results \citep{ade_planck_2016}.

The final catalogue contains $\roughly1.7 \times 10^7$ sources, with $\roughly 15$ million classified as galaxies and $\roughly 2$ million as quasars. This translates to an average number density of more than $600$ sources per sq. deg. in the sky.

\section{Mathematical formalism}
\label{sec:math}

This section describes the summary statistics used to quantify clustering strength. For each statistic, we give a mathematical definition followed by a computational recipe to calculate the clustering strength for given data using that particular statistic. We discuss the auto-clustering of a set of tracers in section~\ref{subsec:mathauto} and the cross-clustering between two sets of tracers in section~\ref{subsec:mathcross}. In section~\ref{subsec:example}, we present an illustrative example that demonstrates the potential of the nearest-neighbour formalism to measure the clustering of a sparse sample of tracers and motivates their application to a BBH-galaxy cross-correlation study.

\subsection{Auto-clustering}
\label{subsec:mathauto}

Consider a set $X$ of $N_X$ discrete, point-like tracers or data points\footnote{We use tracer and data point interchangeably in this paper.}. 
\begin{equation}
    X \equiv \left\{\left(\delta_1, \alpha_1\right), \left(\delta_2, \alpha_2\right), ..., \left(\delta_{N_X}, \alpha_{N_X}\right)\right\} \label{eq:1}
\end{equation}
where $\delta_i$, $\alpha_i$ represents the declination and right ascension of the $i^{\rm th}$ tracer in celestial equatorial (J2000) coordinates. In polar coordinates, the position of the $i^{\rm th}$ tracer is given by 
\begin{align}
    \theta_i &= \delta_i - \pi/2 \label{eq:2a} \tag{2b} \\
    \phi_i &= \alpha_i \label{eq:2b} \tag{2a}
\end{align}

To study clustering, we need a metric for the distance between two points $\left(\delta_1, \alpha_1\right), \left(\delta_2, \alpha_2\right)$ in the sky. A natural choice is the great-circle distance given by $d = \theta$, where $\theta$ is the central angle between the two points on the sphere. Note that we treat the sky as a unit sphere; hence, the radius term that usually multiplies the angle to get the distance is absent. The central angle between $\left(\delta_1, \alpha_1\right), \left(\delta_2, \alpha_2\right)$ is computed using the haversine formula \citep{rios_memoria_1795}
\setcounter{equation}{2}
\begin{equation}
    \text{hav}\left(\theta\right) = \text{hav}\left(\delta_2 - \delta_1\right) + \cos\delta_1\cos\delta_2\text{hav}\left(\alpha_2 - \alpha_1\right) \label{eq:3}
\end{equation}
where $\text{hav}\left(\theta\right) \triangleq \sin^2\left(\theta/2\right)$ is the haversine function.

\subsubsection{Angular Power Spectrum}
\label{subsubsec:twopoint}

The sky positions of the tracers $X$ can be used to define a number density field $n_X(\theta, \phi)$ in the sky, such that 
\begin{equation}
    \int_{\rm All sky}n_X(\theta, \phi)\sin\theta d\theta d\phi = N_X \label{eq:4}
\end{equation}
This can be done numerically, for example, by dividing the sky into equal-area pixels using some pixelation scheme, counting the number of tracers in each pixel, and dividing the number counts by the area of each pixel. Let the average tracer number density in the sky be $\bar{n}_X = \frac{N_X}{4\pi}$. The overdensity field is given by
\begin{equation}
   \delta_X(\theta, \phi) = \frac{n_X(\theta, \phi)}{\bar{n}_X} - 1 \label{eq:5}
\end{equation}
The overdensity field contains all the information about the auto-clustering of the tracer set $A$. By expanding $\delta_X(\theta, \phi)$ in terms of spherical harmonics $Y_{\ell m}(\theta, \phi)$, one can derive the angular power spectrum $\mathcal{C}^{X, X}_{\ell}$, a widely used two-point summary statistic for clustering:
\begin{align}
    &\delta_X(\theta, \phi) = \sum_{\ell m} \alpha^{X}_{\ell m} Y_{\ell m}(\theta, \phi) \label{eq:6} \\
    &\mathcal{C}^{X,X}_{\ell} = \frac{1}{2\ell+1}\sum_{m=-\ell}^{\ell} \lvert\alpha_{\ell m}^{X}\rvert^2 \label{eq:7}
\end{align}
where $\ell$ goes from 0 to $\infty$ and $m$ takes values from $-\ell$ to $\ell$. In practice, the summation is cut off at some $\ell_{\rm max}$ determined by the resolution of the numerical grid on which the field is defined. The power spectrum $C^{X, X}_{\ell}$ at a particular value of $\ell$ quantifies the clustering strength at an angular scale corresponding roughly to $\theta \approx \pi/\ell$. In this study, we use the HEALPix\footnote{\url{https://healpix.sourceforge.io/}} scheme \citep{gorski_healpix_2005} as implemented in the \texttt{python} library \texttt{healpy}\footnote{\url{https://healpy.readthedocs.io/en/latest/}} \citep{zonca_healpy_2019} to compute the overdensity field on a grid of equal-area pixels in the sky. We use the \texttt{healpy}'s \texttt{anafast} routine to compute the power spectra. We choose the default high-$\ell$ cutoff of 3$\times$\texttt{NSIDE} - 1 in our analysis.

An equivalent measure of spatial clustering of a set of tracers is the two-point autocorrelation function, $w^{X, X}(\theta)$, which captures the excess probability of finding two data points separated by an angular distance of $\theta$ in the sky over finding two points drawn from a random (Poisson) distribution in the sky. Mathematically,
\begin{equation}
    w^{X, X}(\theta) = \bigg\langle \delta_{X}(\boldsymbol{\hat{\Omega}}_1)\delta_{X}(\boldsymbol{\hat{\Omega}}_2) \bigg\rangle_{\boldsymbol{\hat{\Omega}}_1 \cdot \boldsymbol{\hat{\Omega}}_2 = \cos\theta} \label{eq:8}
\end{equation}
where $\boldsymbol{\hat{\Omega}}_1$ and $\boldsymbol{\hat{\Omega}}_2$ are unit vectors separated by an angular distance $\theta$ in the sky such that $\boldsymbol{\hat{\Omega}}_1 \cdot \boldsymbol{\hat{\Omega}}_2 = \cos\theta$, and the angular brackets denote an average over all such configurations of $\boldsymbol{\hat{\Omega}}_1$ and $\boldsymbol{\hat{\Omega}}_2$. It can be shown that the angular power spectrum is related to the two-point function of the tracers in the following way:
\begin{equation}
    w^{X, X}(\theta) = \frac{1}{4\pi} \sum_{\ell} \left(1+2\ell\right) \mathcal{C}^{X,X}_{\ell} P_{\ell}\left(\cos\theta\right) \label{eq:9}
\end{equation}
where $P_{\ell}\left(\cos\theta\right)$ denotes the Legendre polynomial of order $\ell$ and argument $\cos{\theta}$. Therefore, the two-point function and the angular power spectrum encode the same physical information about the clustering of the tracer set $X$; the angular power spectrum is a two-point clustering statistic.

\subsubsection{Nearest-neighbour distributions: $k$NN-CDFs}
\label{subsubsec:NN}

The nearest-neighbour distributions as a measure of spatial clustering in 3D were introduced in \cite{banerjee_nearest_2021}. Here, we briefly summarise the idea behind these statistics and extend the mathematical formalism to 2D clustering in the sky using angular coordinates. 

The key idea that motivates the nearest-neighbour clustering framework is as follows: all the physical information about the clustering of a set of discrete tracers is contained in the distribution of their number counts, i.e., the number of tracers enclosed inside a randomly chosen spatial region of a given spatial extent. The spatial regions can have an arbitrary geometrical shape as long as there is a way of assigning a spatial extent to them. Since we are concerned with tracers in the sky, which is represented by the surface of a $3$-sphere, we choose to work with spherical caps of area $A = 2\pi(1-\cos\theta)$ to study clustering at an angular scale $\theta$\footnote{Henceforth, we use $\theta$ and $A$ interchangeably.}. 

Suppose we are given the positions of a set $X$ of discrete tracers. Given the discussion above, the fundamental quantity that quantifies their clustering at spatial scale $\theta$ is the probability $\mathcal{P}_{k|A}$ of finding $k$ data points of $X$ in a randomly placed spherical cap of area $A$ in the sky. $\mathcal{P}_{k|A}$ can be written in terms of a generating function $P(z|A)$ as
\begin{equation}
    \mathcal{P}_{k|A} = \frac{1}{k!} \left[\left(\frac{d}{dz}\right)^kP(z|A)\right]_{z=0} \label{eq:11}
\end{equation}
For the case of spherical caps, it can be shown that the generating function is given by \citep{banerjee_nearest_2021}\footnote{\cite{banerjee_nearest_2021} derived this expression for 3D clustering in cartesian coordinates. The argument is similar for 2D clustering in angular coordinates.}
\begin{align}
    P(z|A) =& \exp\bigg[\sum_{k=1}^{\infty} \frac{\bar{n}_X^k\left(z-1\right)^k}{k!} \notag \\
    & \times \int_{A} ... \int_{A} d\boldsymbol{\hat{\Omega}}_1 ... d\boldsymbol{\hat{\Omega}}_k \omega^{(k)}\left(\boldsymbol{\hat{\Omega}}_1, ...\boldsymbol{\hat{\Omega}}_k\right)\bigg] \label{eq:12}
\end{align}
where $\omega^{(N)}$ are the $N$-point correlation functions of the underlying field of the tracers $X$, with $\omega^{(0)} = 0$ and $\omega^{(1)} = 1$ by definition\footnote{Note that the $N$-point functions are defined analogously to the two-point auto-correlation function defined earlier. In fact, $\omega^{(2)} \equiv w$. However, for $N>2$, $\omega^{(N)}$ are defined as functions of $N$ unit vectors $\left\{\boldsymbol{\hat{\Omega}}_1, ..., \boldsymbol{\hat{\Omega}}_N\right\}$ instead of a single angular separation $\theta$, and the average is performed over all possible configurations preserving the polyhedron formed by the $N$ vectors.}. Equation~\ref{eq:12} shows the connection between the number count distribution and the correlation functions usually employed to measure clustering.

An equivalent measure of clustering is the cumulative distribution of the tracer number counts, which represents the probability $\mathcal{P}_{>k|A}$ of having more than $k$ tracers in a randomly chosen spherical cap of $A$. $\mathcal{P}_{>k|A}$ can also be expressed in terms of a generating function $C(z|A)$ given by \citep{banerjee_nearest_2021}
\begin{align}
    C(z|A) = \frac{1-P(z|A)}{1-z} \label{eq:13}
\end{align}
such that
\begin{equation}
    \mathcal{P}_{>k|A} = \frac{1}{k!} \left[\left(\frac{d}{dz}\right)^kC(z|A)\right]_{z=0} \label{eq:14}
\end{equation} 
From equations~\ref{eq:12} to~\ref{eq:14}, it is not clear how to compute these distributions without first computing all higher-order correlation functions. Following \cite{banerjee_nearest_2021}, we now discuss another interpretation of the cumulative count distributions that allows us to compute $\mathcal{P}_{>k|A}$ directly from the positions of the tracers. The count distributions $\mathcal{P}_{k|A}$ can then be calculated trivially using $\mathcal{P}_{k|A}=\mathcal{P}_{>k-1|A}-\mathcal{P}_{>k|A}$.

Consider a set of $N_r$ area-filling, randomly distributed query points in the sky, such that $N_r \gg N_X$. Each query point will have a data point in $X$ that is nearest to it, a data point that is second-nearest to it and so on. The distributions of the distances to these neighbouring data points, over all query points in the sky, are directly connected to the count distributions discussed above. We argue that the cumulative distribution function (CDF) of the distances from the query points to the $k$-nearest-neighbour data point, or $k$NN-CDF, is precisely equal to $\mathcal{P}_{>k-1|A}$. 

To understand this connection, let us examine the case of $k = 1$. Consider $N_r$ spherical caps of area $A = 2\pi\left(1-\cos\theta\right)$, the centres of which are distributed randomly in the sky. The fraction of such caps enclosing at least 1 data point is equal to that of cap centres with the angular distance to their nearest neighbour less than $\theta$. The nearest-neighbour CDF at angular scale $\theta$ is the precise measure of the fraction of query points (equivalent to centres of the spherical caps) for which the nearest data point is at a distance less than $\theta$. This argument can be easily generalised if we consider the $k$-nearest-neighbour instead of the first nearest-neighbour. Therefore, we conclude
\begin{equation}
    \mathcal{P}_{>k-1|A} = \text{CDF}_{k\rm NN}(\theta) \label{eq:15}
\end{equation} 
In practice, the $k$NN-CDFs are simple to calculate in a computationally efficient manner. We start by creating a HEALPix grid of query points with a sufficiently high value of \texttt{NSIDE}, such that the resolution of the query grid is much finer than the smallest angular scale at which we want to study spatial clustering. As noted in \cite{banerjee_nearest_2021}, placing query points on a finely spaced grid gives the same results as randomly distributed query points, as long as the grid separation is much smaller than the mean interparticle separation of the data. Next, we compute the distances to the $k$-nearest-neighbour data point of each query point. The nearest-neighbour search is carried out very efficiently by constructing a Ball tree structure \citep{Omohundro2009FiveBC} on the data points. Once a tree is built, it can be used to calculate the distances to the first $k$ neighbouring data points for all query points simultaneously. Sorting the computed distances for each neighbour index $k$ immediately gives the empirical CDF of the $k$-nearest-neighbour distances over a range of spatial scales. The empirical CDF converges to the true $k$NN-CDF in the limit of a large number of query points. In this study, we utilise the \texttt{sklearn.neighbors.BallTree} routine from the library \texttt{scikit-learn}\footnote{\url{https://scikit-learn.org/}} \citep{pedregosa_scikit-learn_2012} with the haversine distance metric for our purposes.

It is evident from equation~\ref{eq:12} that the count distributions $\mathcal{P}_{>k|A}$, and hence the $k$NN-CDFs, are formally sensitive to integrals of all $N$-point correlation functions of the underlying tracer field. This makes these summary statistics extremely powerful probes of clustering on small spatial scales where the higher-order correlation functions contribute significantly. We refer the interested reader to \cite{banerjee_nearest_2021} for a detailed study of the gain in clustering measurements as well as cosmological constraints achieved using the $k$NN-CDFs over two-point clustering statistics.

\subsection{Cross-clustering}
\label{subsec:mathcross}

\begin{figure}
    \centering
    \includegraphics[width=\columnwidth]{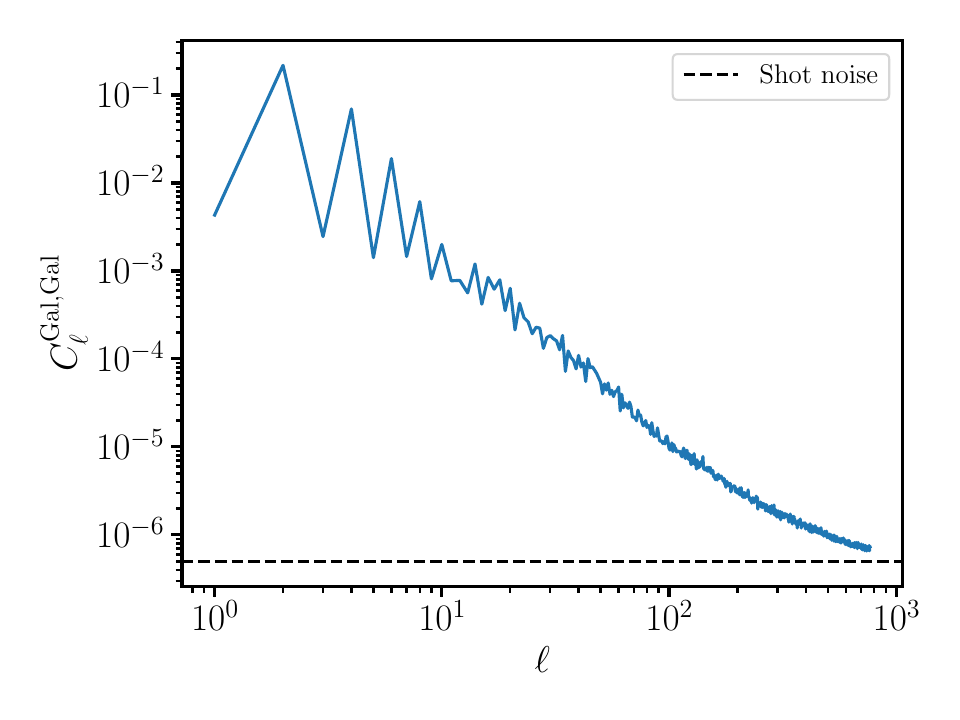}
    \caption{The angular power spectrum for the WSC sources computed using \texttt{healpy}'s \texttt{anafast} routine, with the shot (Poisson sampling) noise plotted for reference. The power spectrum is well above the shot noise up to an $\ell_{\rm max} \roughly 10^3$, corresponding to spatial scales much smaller than those considered in this analysis. Therefore, the treatment of the WSC sources as a continuous field is a reasonable approximation.}
    \label{fig:5}
\end{figure}

There are two possible approaches to quantify the cross-clustering between the BBHs and the WSC catalogue:
\begin{enumerate}
    \item directly cross-correlate the WSC source positions with the BBH positions
    \item cross-correlate fluctuations in the WSC source number density field and the BBH positions.
\end{enumerate}
The number of sources in the WSC catalogue is a few orders of magnitude larger than the number of BBH events under consideration, making a tracer-tracer cross-correlation difficult to compute using the $k$NN formalism. Since the source number density is large enough, on the scales that we will consider in the analysis, the WSC catalogue can be effectively treated as a `continuous' field. Furthermore, as shown in figure~\ref{fig:5}, the Poisson sampling noise, or shot noise, is subdominant to the angular power spectrum at all scales of interest. This means that the underlying galaxy density field is well-sampled by WSC source positions. Therefore we adopt approach (ii) in this paper and compute the BBH-Galaxy cross-clustering using a tracer-field cross-correlation formalism.  

We now describe the summary statistics for measuring the spatial cross-correlation between a set of discrete tracers, $X$, and a continuous field $\delta_{Y}$. We only consider continuous fields in the form of dimensionless fluctuations in a quantity, i.e., fields that are bounded below by -1 and average to 0. The continuous field could be an overdensity field derived from another sample of tracers, such as the galaxies relevant for this study, or an actual continuum field like the CMB temperature fluctuations.

\subsubsection{Cross Angular Power Spectrum}
\label{subsubsec:twopointcross}
As discussed in section~\ref{subsubsec:twopoint}, we can define an overdensity field $\delta_X$ for the discrete tracers $X$, and expand both $\delta_X$ and $\delta_Y$ in spherical harmonics
\begin{align*}
    \delta_X(\theta, \phi) & = \sum_{\ell m} \alpha^{X}_{\ell m} Y_{\ell m}(\theta, \phi) \\
    \delta_Y(\theta, \phi) & = \sum_{\ell m} \alpha^{Y}_{\ell m} Y_{\ell m}(\theta, \phi)
\end{align*}
The cross angular power spectrum between $X$ and $Y$ is defined as
\begin{equation}
    \mathcal{C}^{X,Y}_{\ell} = \frac{1}{2\ell+1}\sum_{m=-\ell}^{\ell} \left\{\alpha^{X}_{\ell m}\right\}^*\alpha^{Y}_{\ell m} \label{eq:16}
\end{equation}
We compute the cross angular power spectrum in a similar manner to the auto angular power spectrum, using \texttt{healpy}'s \texttt{anafast} routine.

\subsubsection{Nearest-neighbour distributions}
\label{subsubsec:NNcross}

\cite{banerjee_tracer-field_2023} generalised the $k$NN formalism to study tracer-field cross-correlations in 3D using the nearest-neighbour distributions. Here, we summarise the main points in the context of 2D angular clustering. 

To perform nearest-neighbour measurements on a continuous field, we must define the continuum version of the $k$NN-CDFs. In other words, we must investigate the behaviour of the $k$NN-CDFs of a set of discrete tracers of the field $\delta_Y$ as their average number density $\bar{n}_Y$ tends to infinity. If the tracers represent a local Poisson process on the field $\delta_Y$, \cite{banerjee_tracer-field_2023} showed that the continuum version of the $k$NN measurements at a spatial scale $\theta$ are thresholded evaluations of the CDF of the smoothed continuous field (see also \cite{10.1093/mnras/stad3118} for an application of the CDF formalism to the lensing convergence fields from the first three years of the Dark Energy Survey, where a similar approach was taken). Mathematically,
\begin{equation}
    \mathcal{P}_{\geq k|A} \to \mathcal{P}_{> \delta^*}(\theta) = \int_{\delta^*}^{\infty}\phi(\delta^{\theta}_{Y})d\delta^{\theta}_{Y} = 1 - \text{CDF}(\delta^*) \label{eq:22}
\end{equation}
where $\delta^{\theta}_{Y}$ represents the continuous field smoothed on the scale $\theta$, $\phi(\delta^{\theta}_{Y})$ represents the probability density function (PDF) of the smoothed field. The nearest-neighbour index $k$ for discrete data maps to a threshold density $\delta^*$ on the smoothed continuous field, such that $\bar{n}_Y A \left(1+\delta^*\right) = k$\footnote{Note that at fixed $k$, $\delta^*$ is also a function of $\theta$, as evident from its definition}.

Now that we have a continuum version of the $k$NN-CDFs, we discuss the characterisation of spatial cross-correlations using the $k$NN formalism. Following \cite{banerjee_tracer-field_2023}, we take the joint probability $\mathcal{P}_{\geq k, >\delta^*}(\theta)$ of finding at least $k$ tracers and the smoothed continuous field $\delta^{\theta}_Y$ to cross threshold $\delta^*$ in spherical caps of angular radius $\theta$ as a measure of the spatial cross-correlations between tracers and a continuous field. Assuming that the tracers $X$ represent a local Poisson process on the overdensity field $\delta_X$, we have, similar to \cite{banerjee_tracer-field_2023},
\begin{equation}
    \mathcal{P}_{k, >\delta^*}(\theta) = \int_{\delta^*}^{\infty} \frac{\left[\lambda(\delta^{\theta}_X)\right]^k}{k!}e^{-\lambda(\delta^{\theta}_X)} \phi(\delta^{\theta}_{X}, \delta^{\theta}_{Y})d\delta^{\theta}_{X}d\delta^{\theta}_{Y} \label{eq:23}
\end{equation}
where $\phi(\delta^{\theta}_{X}, \delta^{\theta}_{Y})$ is the joint probability distribution of the two fields when smoothed on angular scale $\theta$. The quantity of our interest, $\mathcal{P}_{\geq k, >\delta^*}$, can be written in terms of $\mathcal{P}_{k, >\delta^*}$ as
\begin{equation}
    \mathcal{P}_{\geq k, >\delta^*} = \mathcal{P}_{>\delta^*} - \sum_{j<k}\mathcal{P}_{j, >\delta^*} \label{eq:24}
\end{equation}
Suppose the tracers $X$ are completely uncorrelated and statistically independent of the continuous field $\delta_Y$. In that case, the joint distribution function can be factored into a product of the individual PDFs of the smoothed fields, i.e., $\phi(\delta^{\theta}_{X}, \delta^{\theta}_{Y}) \propto \phi(\delta^{\theta}_{X})$ $\phi(\delta^{\theta}_{Y})$. In this case, we have \citep{banerjee_tracer-field_2023} $\mathcal{P}_{\geq k, >\delta^*} = \mathcal{P}_{\geq k} \times \mathcal{P}_{>\delta^*}$. This fact provides a convenient way to define a summary statistic that measures the \textit{excess cross-correlation} between the tracers and the continuous field
\begin{equation}
    \psi_{k, \delta*} \triangleq \mathcal{P}_{\geq k, >\delta^*}/\left(\mathcal{P}_{\geq k} \times \mathcal{P}_{>\delta^*}\right) \label{eq:25}
\end{equation}
This is a useful quantity to measure, as a positive (negative) measurement for $\psi_{k, \delta*} - 1$ would indicate that the tracer $X$ is correlated (anti-correlated) with the field $\delta_Y$, while $\psi_{k, \delta*} - 1 = 0$ would indicate that the there is no spatial cross-correlation between them.

All the physical information about the spatial cross-correlation between fluctuations in the sky distribution of tracer $X$ and the field $Y$ is contained in the joint distribution $\phi(\delta^{\theta}_{X}, \delta^{\theta}_{Y})$. Therefore, it is clear from equations~\ref{eq:23} to~\ref{eq:25} that the excess cross-correlation, as defined using the nearest-neighbour distributions, will be sensitive not just to the linear or Gaussian correlations in the density fluctuations of the tracers and the continuous field, but to correlations in fluctuations at all orders (See \cite{banerjee_tracer-field_2023} for a detailed demonstration). Therefore, the $k$NN formalism provides a powerful way to characterise cosmological cross-correlations.

The joint probability distributions $\mathcal{P}_{\geq k, >\delta^*}$ and excess cross-correlations $\psi_{k, \delta*}$ are simple to compute numerically. We follow the procedure laid out in \cite{banerjee_tracer-field_2023}
\begin{enumerate}
    \item Create a set of area-filling query points by creating a finely-spaced HEALPix grid in the sky, such that the number of pixels \texttt{$N_{\rm{pix}}$} is far greater than the number of data points.
    \item Build a Ball tree from the set of tracer positions and estimate the angular distances to the $k$-nearest neighbour data points from each query point. For each $k$, sort the distances to produce the empirical $k$NN-CDF over a range of angular scales $\theta$. In the limit of large \texttt{$N_{\rm{pix}}$}, the empirical CDF approaches $\mathcal{P}_{\geq k}$.
    \item Smooth the continuous field $\delta_Y$ on an angular scale $\theta$ using a top-hat filter. The smoothing is done in harmonic space using the $\{\alpha^{Y}_{\ell m}\}$ of the field, computed via spherical harmonic transforms, to speed up the computation time (see appendix~\ref{sec:D} for details). Interpolate the smoothed field on the query grid defined in step (i).
    \item For a given $k$ and threshold $\delta^*$, compute the fraction of query points for which the $k^{\rm th}$ nearest-neighbour lies at an angular distance less than $\theta$ and the smoothed field, interpolated to that grid point, exceeds $\delta^*$. In the limit of large \texttt{$N_{\rm{pix}}$}, this fraction approaches $\mathcal{P}_{\geq k, >\delta^*}$.
    \item Compute the fraction of query points for which the smoothed field, interpolated to the grid point, exceeds $\delta^*$. In the limit of large \texttt{$N_{\rm{pix}}$}, this fraction approaches $\mathcal{P}_{>\delta^*}$.
    \item From the quantities calculated above, compute the excess cross-correlation using equation~\ref{eq:25}.
    \item Repeat steps (iii) to (vi) for different values of the angular scale $\theta$.
\end{enumerate}

In this study, we choose the \textit{constant percentile} threshold described in \cite{banerjee_tracer-field_2023} to define the threshold value $\delta^*$ for the continuous field. Specifically, we set $\delta^* = \delta_{75}$, the value of the $75^{\rm th}$ percentile of $\delta^{\theta}_Y$. This choice implies that $\mathcal{P}_{>\delta^*} = 0.25$ irrespective of the smoothing scale $\theta$.

\subsection{Example application of the tracer-field formalism}
\label{subsec:example}

As discussed in the previous section, the nearest-neighbour distributions are sensitive to all higher-order cross-correlation functions of the discrete tracers and the continuous field. Consequently, nearest-neighbour measurements are expected to measure a stronger clustering signal than the two-point summary statistics at small angular scales where the underlying fields are non-Gaussian, and these higher-order correlation functions cannot be neglected. In this section, we demonstrate the gains in clustering power obtained using the $k$NN tracer-field formalism compared to the angular power spectrum using an illustrative example. 

We create a set of discrete tracers by picking the centres of $36$ pixels in the WSC footprint\footnote{This is chosen to match the average number of mock BBHs in the WSC footprint, see figure~\ref{fig:8} for details.} that have the highest number density of sources in the sky, and compute their spatial cross-correlations with the full WSC overdensity field using both the nearest-neighbour measurements and the angular power spectrum. We further assume that we know the locations of these tracers perfectly. To estimate the cosmic variance associated with the clustering measurements, we also compute these clustering statistics for 100 realisations of $36$ randomly chosen points in the sky. We compute the cross-correlations on angular distance scales from $\roughly 1^{\circ}$ to $\roughly 35^{\circ}$, equivalent to $\ell=6$ to $\ell=180$. This choice ensures that 
\begin{enumerate}
    \item we have sufficient sampling in the nonlinear regime, where the nearest-neighbour distributions can capture information not accessible through two-point statistics
    \item the measurements are not affected by the lack of sampling towards the right tail of the auto $k$NN-CDF (see \cite{banerjee_nearest_2021} for a discussion)
\end{enumerate} 
These angular distances correspond to projected transverse distance scales of $\roughly 15$ to $\roughly 400$ Mpc for a median redshift of $\roughly 0.2$ for the WSC catalogue.

\begin{figure*}
    \centering
    \begin{multicols}{2}
        \includegraphics[width=\linewidth]{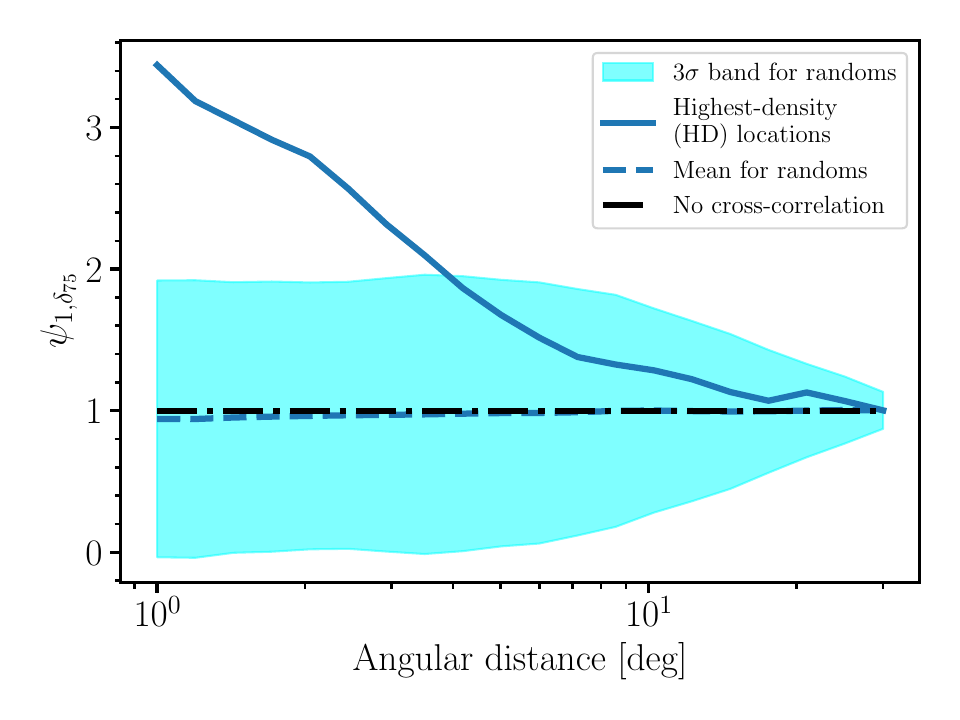}\par
        \includegraphics[width=\linewidth]{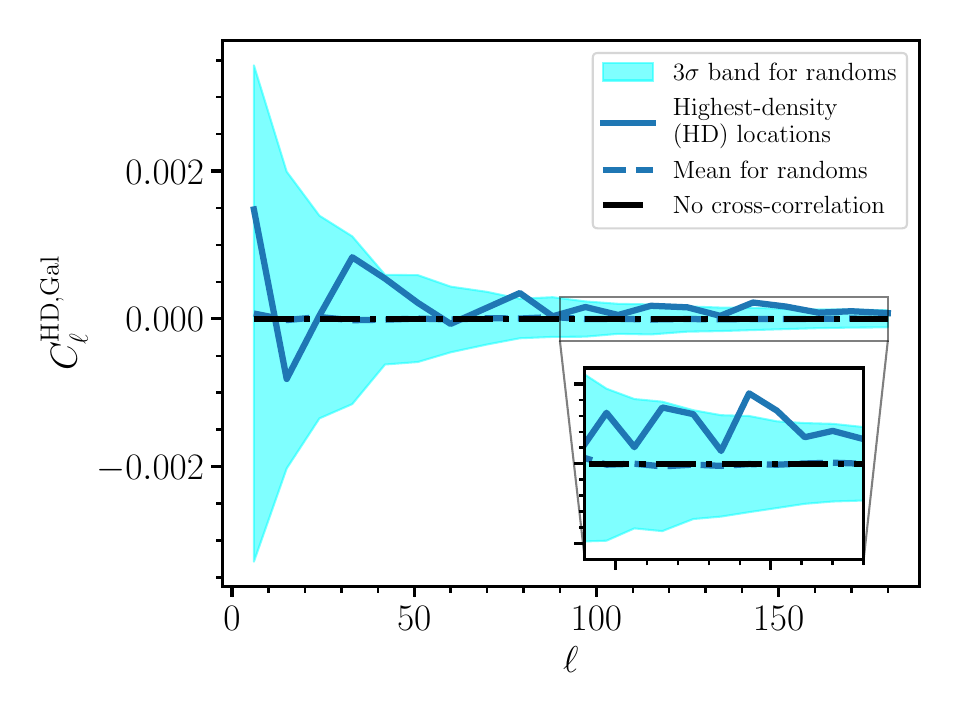}\par 
    \end{multicols}
    \caption{The excess cross-correlation between 36 locations with the highest density of WSC sources and the WSC overdensity map as measured by the first nearest-neighbour distribution (left) and the cross angular power spectrum (right). The solid lines represent the measurement on the tracers, the shaded band represents the 3$\sigma$ variance in the same measurement performed on randomly chosen points in the sky, and the dash-dot line represents the expected value in the absence of cross-correlation. The inset on the right panel shows a zoomed-in view of the power spectrum at smaller scales.}
    \label{fig:ex1}
\end{figure*}

The results for the first nearest-neighbour distribution are shown in the left panel of figure~\ref{fig:ex1}, and those for the power spectrum are shown in the right panel. The inset in the right panel presents a zoomed-in version of the full subplot focusing on the smaller scales. In each plot, the solid line shows the excess cross-correlation between the highest-density locations, the shaded band represents the cosmic variance, and the dash-dot line represents the expected value in the absence of cross-correlation.

The excess cross-correlation, as measured by the first nearest-neighbour distribution, lies well outside the shaded region representing the $3\sigma$ cosmic variance in the randoms at all angular scales smaller than $\roughly 4^{\circ}$. In contrast, the power spectrum fluctuates within the shaded band, barely crossing the detection threshold on one or two bins. Thus, figure~\ref{fig:ex1} demonstrates that the nearest-neighbour measurements can capture a statistically significant clustering signal on small, nonlinear scales for a well-localised sample of rare, highly biased tracers, whereas the power spectrum can not. We would like to draw the attention of the reader to the following important implication of this finding: \textit{with a set of $<50$ well-localised events, the nearest-neighbour measurements on small spatial scales are statistically robust enough to investigate whether BBHs reside in highly biased environments in the universe}. It should be noted, however, that on large scales, the nearest-neighbour distributions do not perform any better than the power spectrum, and there is no detectable signal in either statistic. This is because, on large scales, the galaxy density field is well approximated by a Gaussian random field, and the higher-order correlation functions are negligible.

\section{Methods}
\label{sec:methods}

In this section, we describe the procedure to quantify the clustering of the BBHs and their spatial cross-correlations with the WSC sources. We address this problem using a hypothesis-testing approach: We consider a null hypothesis, which proposes that there is no statistical significance for a clustering signal in the data, and attempt to rule it out by investigating the likelihood of reproducing the observed data, assuming the null hypothesis to be true. We describe our null hypothesis in section~\ref{subsec:NH}. To test the null hypothesis, we require a control dataset consistent with its premise. We already described the procedure for creating a catalogue of unclustered mock BBHs in section~\ref{subsec:mockdata}. This mock catalogue automatically serves as a control set to test the null hypothesis. In section~\ref{subsec:application}, we describe how we apply the mathematical formalism motivated in section~\ref{sec:math} to compute clustering statistics for the specific data considered in this study. We discuss the method to calculate the statistical significance of the clustering signal in section~\ref{subsec:significance}. Finally, we describe the angular distance scales used in our analysis in section~\ref{subsec:angscale}.

\subsection{Null Hypothesis}
\label{subsec:NH}

Our null hypothesis is as follows
\begin{quote}
    \textit{The BBHs currently detected by the LVK collaboration are spatially unclustered, distributed uniformly (isotropically) in the sky and are not spatially correlated with other tracers of the large-scale structure of the universe, such as galaxies and quasars.}
\end{quote}
Any dataset consistent with this hypothesis would not contain a statistically significant clustering signal. In the following sections, we discuss the process of testing the null hypothesis using the observed and mock BBH catalogues described in section~\ref{sec:data}.

\subsection{Application of clustering formalism to data}
\label{subsec:application}

The formalism to study clustering developed in section~\ref{sec:math} is applicable to discrete tracers that can be treated as point objects in the sky, i.e., objects that can be localised to a single sky position $(\delta, \alpha)$. However, as shown in section~\ref{sec:data}, due to the limited resolving power of gravitational wave detectors, we can only assign each event with a probability distribution over an extended region in the sky (see figure~\ref{fig:1} for example). Moreover, the computational procedure outlined in section~\ref{sec:math} gives unbiased measurements for the tracer-field cross-correlations only when the continuous field is defined on the entire sky. However, as discussed in section~\ref{sec:data}, certain regions in the sky do not have reliable galaxy data. Hence, we can not define the galaxy overdensity field there. In this section, we discuss our strategy to deal with these challenges, namely the uncertain sky localisation of the BBHs and the complicating effects due to the presence of the WSC mask. 

\subsubsection{Strategy to deal with the uncertainty in BBH sky localisations}
\label{subsubsec:str1}

For each BBH, we have a probability distribution in the sky. Since we need a single location for the BBHs, one possible approach is to assign each BBH the most probable position in its sky localisation area, i.e., the position where the probability distribution is maximised. However, the sky posteriors of many events show signs of bimodality, and assigning a single representative location to them is problematic. Furthermore, in reducing a probability distribution to a single point, we lose a lot of information contained in the shapes of the posteriors. For example, utilising the full sky distribution can also allow us to characterise the measurement errors on the clustering strength. Therefore, we adopt a different strategy in this work, which is as follows
\begin{enumerate}
    \item for each BBH, draw an (RA, Dec) pair from the sky location posterior
    \item using the drawn samples as the `true' locations of the events, compute the auto-clustering statistics as defined in section~\ref{subsec:mathauto}
    \item repeat steps (i) and (ii) for 1000 draws from the posteriors
\end{enumerate}
The average over the 1000 draws gives the estimated value of the clustering strength, while the variance over the 1000 draws gives the estimated measurement error due to the uncertainty in the sky localisation of the BBHs. Therefore, our strategy naturally preserves the information present in the full sky distribution of each BBH while giving us an estimate of the measurement errors on the clustering strength. \cite{vijaykumar_probing_2023} took a similar approach in a recent clustering analysis with forecast BBH data for the third generation of gravitational wave detectors.

\subsubsection{Strategy to deal with complications due to the WSC Mask}
\label{subsubsec:str2}

\begin{figure}
    \centering
    \includegraphics[width=\columnwidth]{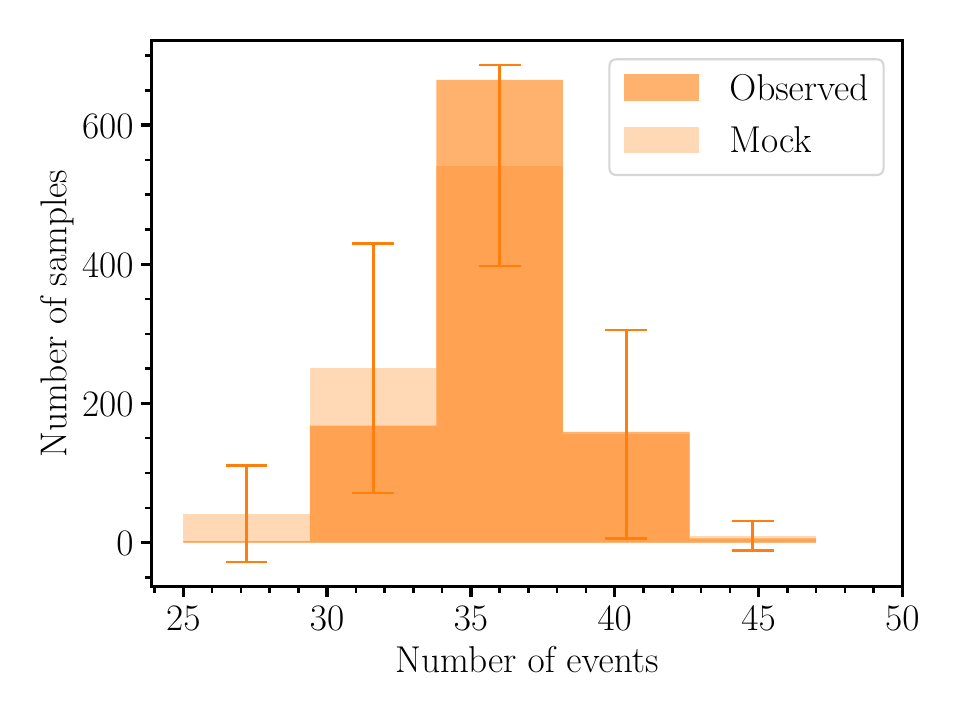}
    \caption{Distribution of the number of unmasked events over 1000 samples drawn from the sky distributions of the observed (bold histogram) and mock (light histogram) BBHs. The error bars on the mock histogram show the $1\sigma$ variance over 135 realisations of the mock catalogue. There is an excellent match between the two distributions within error bars. Therefore, any systematics that arise due to differing number densities between different samples would affect the clustering of the observed and mock BBHs equally.}
    \label{fig:8}
\end{figure}

Since we do not have reliable data in regions outside the WSC mask, the galaxy overdensity field is ill-defined there. How do we compute cross-correlations with the BBHs in this scenario? One possible approach is to assign a $\delta_{\rm Gal} = 0$ to all pixels in the masked region since that is the expected average value of an overdensity field. Although this would not bias the results since the BBHs outside the mask would not contribute to the cross-correlation signal, it would unnecessarily add to the noise budget. Instead, we take the approach of removing the BBH events that lie outside the mask. However, since the BBHs are not perfectly localised, we must be careful when handling events whose sky localisation areas are partly inside the mask. We follow the following strategy:
\begin{enumerate}
    \item Draw 1000 samples of (RA, Dec) pairs from the sky location posteriors of each of the BBHs
    \item for each sample, remove the sky locations that lie outside the WSC mask
\end{enumerate}
By keeping the posterior samples for events which are partly outside the mask, our method not only preserves the information in the sky distribution of the BBHs but also leads to more number of BBHs contributing to the analysis than simply removing all BBHs whose localisation area intersects the mask would. However, this process leads to different tracers in each sample. Since the $k$NN-CDFs are highly sensitive to the number density of the tracers (see equation~\ref{eq:12}), care needs to be taken to ensure that averaging the CDFs over samples with different number densities does not lead to any issues. An important check is whether the distribution of the number of events inside the mask over the 1000 samples for the mock catalogue generated in section~\ref{subsec:mockdata} is statistically similar to that of the data. Figure~\ref{fig:8} shows that, indeed, that is the case, and hence any systematics that arise due to this would affect the clustering of the observed and mock BBHs equally.

We now have positions for the BBHs, from which we can compute the overdensity fields needed to compute the cross angular power spectrum. Since there is no data outside the mask, the value of the overdensity field computed there would be artificially low (negative). To account for this, we set the BBH overdensity fields outside the mask to zero before calculating their spherical transforms. 

To compute the nearest-neighbour measurements of the BBH-Galaxy excess cross-correlation, we need to smooth the galaxy density field on various spatial scales. To minimise the effects due to the mask, we set the field outside the mask to zero before smoothing. We restrict the query points to inside the mask. This is needed because the query points outside the mask would have artificially large nearest-neighbour distances and would skew the measured distributions. We also remove all query points within a certain angular distance from the mask boundaries to ensure that the smoothed density field interpolated at the query points is not affected by spurious contributions from the regions outside the mask. \cite{10.1093/mnras/stac1551} followed a similar procedure to analyse the spatial clustering of SDSS clusters using $k$NN-CDFS in a recent study. In practice, we observe that a threshold distance of roughly half the maximum angular scale used in the analysis leads to an unbiased measurement of the excess cross-correlation. Note that these steps are not taken when computing the BBH auto-clustering since we have BBH data on the entire sky.

\subsection{Statistical significance}
\label{subsec:significance}

In this section, we describe the way we compute the statistical significance of the clustering measurements once the summary statistics have been computed for the observed and mock BBHs. Consider a summary statistic as a function of angular scale, $S(\theta)$, which could be the angular power spectrum, the two-point function or the nearest-neighbour distribution, either as a measure of the BBH auto-correlation or BBH-Galaxy cross-correlation. Let the scales considered in the analysis be $\{\theta_1, ..., \theta_p\}$. We define the \textit{data vector} $D_a$ as the summary statistic $S$ evaluated on the observed BBH catalogue at angular scale $\theta_a$. Similarly, we define a \textit{mock vector} $M^i_b$ as $S(\theta_b)$ evaluated on the $i^{\rm th}$ realisation of the mock BBH catalogue for each of the $n$ realisations. Note that $S(\theta)$ represents the summary statistic already averaged over 1000 samples drawn from the sky distribution of the BBHs, as prescribed in section~\ref{subsubsec:str1}. To characterise the noise properties of the measurement, we compute the \textit{covariance matrix}
\begin{equation}
    \boldsymbol{\Sigma}^{'}_{ab} = \bigg\langle \left(M^i_a - \langle M_a \rangle \right) \left(M^i_b - \langle M_b \rangle \right) \bigg\rangle \label{eq:26}
\end{equation}
where the angular brackets denote an average over the $n$ realisations of the mock catalogue. The covariance matrix is, by definition, a $p \times p$ matrix. The object that is relevant for the statistical calculations is the inverse of the covariance matrix, which is multiplied by the Hartlap correction factor \citep{ refId0} to get an unbiased estimate
\begin{equation}
    \boldsymbol{\Sigma}^{-1} = \frac{n-p-2}{n-1} \left(\boldsymbol{\Sigma}^{'}\right)^{-1} \label{eq:27}
\end{equation}
Once we have the corrected inverse covariance matrix, we characterise the signal-to-noise for clustering using the $\chi^2$ statistic. For the observed BBHs and each realisation of the mock BBHs, we define the $\chi^2$ value as
\begin{align}
    \chi^2_{D} &= \big(\boldsymbol{D} - \langle \boldsymbol{M} \rangle \big)^T \boldsymbol{\Sigma}^{-1} \big(\boldsymbol{D} - \langle \boldsymbol{M} \rangle \big) \label{eq:28a} \\
    \chi^2_{M^i} &= \big(\boldsymbol{M}^i - \langle \boldsymbol{M} \rangle \big)^T \boldsymbol{\Sigma}^{-1} \big(\boldsymbol{M}^i - \langle \boldsymbol{M} \rangle \big) \label{eq:28b}
\end{align}
where $\boldsymbol{D} \triangleq \{D_1, D_2, ..., D_p\}$ and $\boldsymbol{M}^i \triangleq \{M^i_1, M^i_2, ..., M^i_p\}$. The distribution of $\chi^2_{M^i}$ (henceforth the null distribution) represents the signal-to-noise expected from data consistent with the null hypothesis, and $\chi^2_{D}$ represents the signal-to-noise measured from the data. A larger value for $\chi^2_{D}$ relative to the null distribution implies a stronger statistical significance of the clustering signal. 

From the null distribution and the measured signal-to-noise, we compute the $p$-value, or probability of reproducing the observations assuming the null hypothesis is true, by estimating the area enclosed under the (normalised) null distribution curve after it crosses the measured signal-to-noise. In practice, this can be estimated by counting the fraction of mock realisations with $\chi^2_{M^i} > \chi^2_{D}$. If the signal is strong enough that none of the mock realisations have a larger $\chi^2$ than the data, then one must fit a $\chi^2$ distribution to the null distribution to compute the $p$-value. The null hypothesis is ruled out if the $p$-value is smaller than a chosen detection threshold.

\subsection{Angular scales}
\label{subsec:angscale}

Similar to section~\ref{subsec:example}, we choose $10$ log-spaced angular bins from $\theta \roughly 1^{\circ}$ to $\theta \roughly 35^{\circ}$ for the clustering analysis, which leads to a Hartlap factor of $0.92$. For the angular power spectrum, we choose $10$ linearly-spaced bins between $\ell = 6$ and $\ell = 180$. 

For computing the overdensity fields and the query points for the nearest-neighbour measurements, we use an \texttt{NSIDE} = 256 HEALPix grid with $\roughly 7.8 \times 10^5$ pixels and an angular resolution of $\roughly 0.22^{\circ}$. As required for the nearest-neighbour analysis, the number of query pixels is much larger than the number of data points, and the query grid has sufficient resolution to sample the smallest spatial scales analysed. 

As discussed in section~\ref{subsubsec:str2}, we remove all query points within $20^{\circ}$ of the WSC mask boundaries for computing nearest-neighbour excess cross-correlation to avoid any biases due to the presence of the WSC mask.

\section{Results}
\label{sec:results}

In this section, we present the results of our clustering analysis, with section~\ref{subsec:resang} devoted to the auto and cross angular power spectrum and section~\ref{subsec:resknn} to the nearest-neighbour measurements. We discuss the implications of our findings in section~\ref{subsec:disc}.

\subsection{Angular power spectrum}
\label{subsec:resang}

\begin{figure}
    \centering
    \includegraphics[width=\columnwidth]{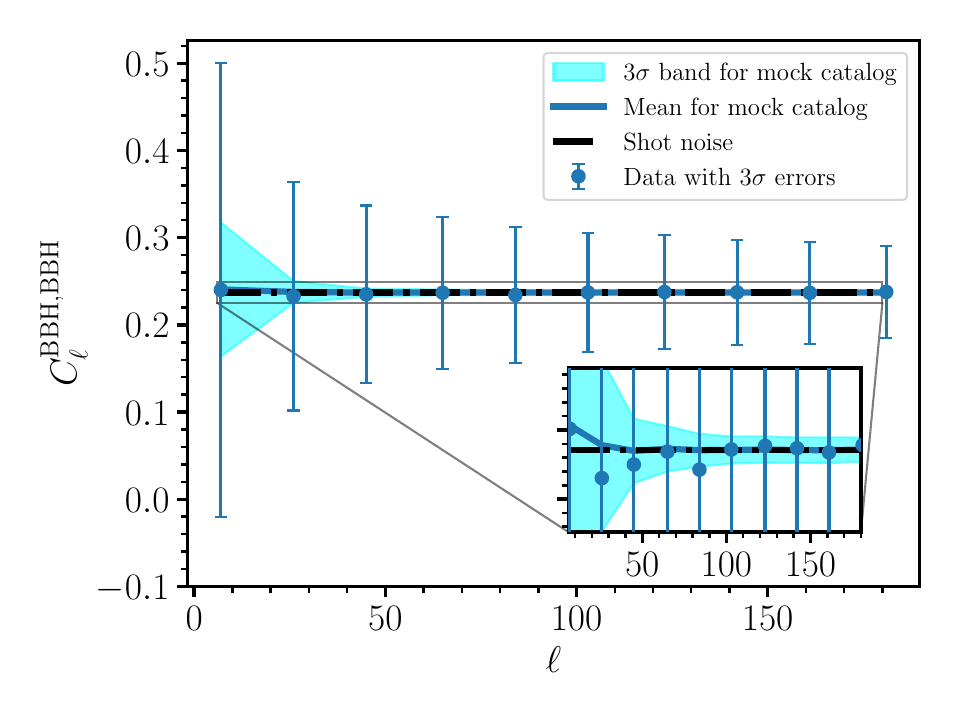}
    \caption{Results of auto-clustering analysis conducted using the angular power spectrum as the summary statistic, with a zoomed-in view provided in the inset for better visibility. Filled circles represent the measured angular power spectrum of the observed BBHs, with error bars representing variance across 1000 samples drawn from the BBH skymaps. The bold line represents the angular power spectrum averaged over 135 realisations of the mock BBH catalogue. The shaded band, which shows the variance of the power spectrum across realisations, represents the cosmic variance. The dash-dot line shows the shot noise. All errors are displayed at the $3\sigma$ level. It is evident from the figure that with the present number of BBH detections, the angular power spectrum is shot noise-dominated; visually, there are no signs of a clustering signal.}
    \label{fig:9}
\end{figure}

Figure~\ref{fig:9} shows the angular power spectrum of the BBHs. Filled circles represent the power spectrum of the observed BBHs, and the error bars are measurement errors, defined as 3 times the standard deviation across 1000 samples drawn from the BBH skymaps. The bold line and shaded band show the mean angular power spectrum and $3\sigma$ variation around the mean for 135 realisations of the mock BBH catalogue. The dash-dot line represents the Poisson sampling noise (shot noise) corresponding to the number density of the BBH sample, which is equal to $1/(\bar{n}_{\rm BBH})$. Since the error bars make it difficult to visualise the shaded band, we plot a zoomed-in version of the full figure in the inset.

The mean power spectrum for the mock catalogue is very close to the shot noise, as should be the case for a dataset consistent with the null hypothesis. The variance over realisations of the mock catalogue gives an estimate of the cosmic variance expected in the power spectrum if the null hypothesis holds. It is evident from the figure that with the present number of BBH detections, the angular power spectrum is shot noise-dominated and cannot capture a clustering signal. As can be seen from the figure, the measurement errors on the data are extremely large and exceed the cosmic variance. This is a consequence of the significant uncertainties in the sky-localisation of the BBHs.

\begin{figure}
    \centering
    \includegraphics[width=\columnwidth]{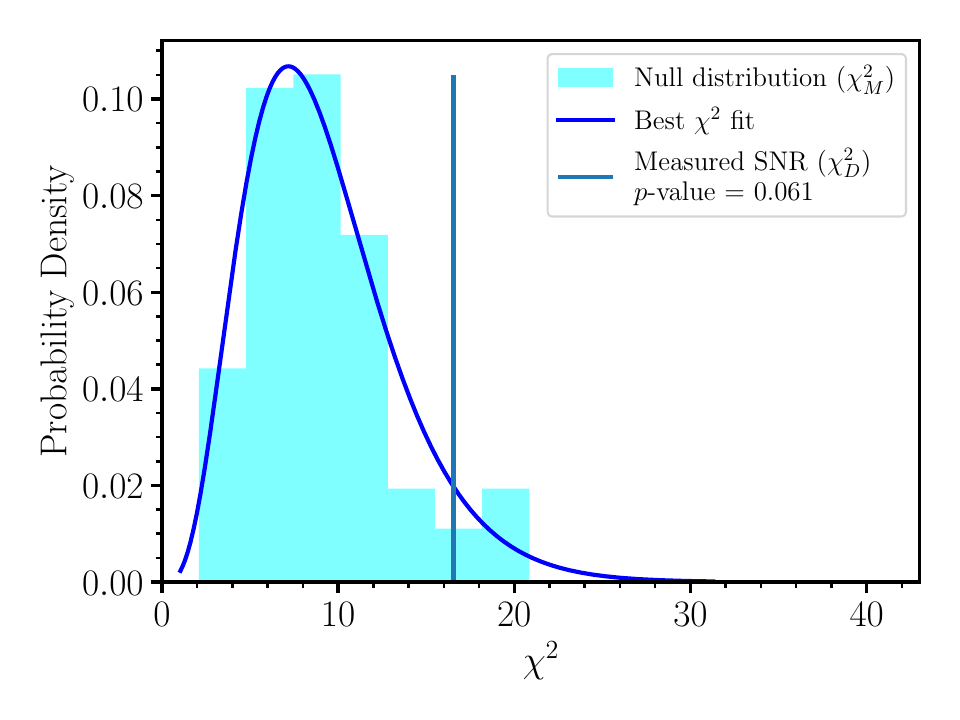}
    \caption{Results of the statistical significance test for the auto angular power spectrum of the BBHs. The histogram represents the distribution of $\chi^2$ values over 135 realisations of the mock catalogue, and the curve enveloping it represents the best-fit $\chi^2$ distribution. The vertical line represents the measured $\chi^2$ value for the data. The data is consistent with the null hypothesis at a $p$-value of 0.061, calculated using the CDF of the best $\chi^2$ fit. There is no evidence for a statistically significant clustering signal in the present data.}
    \label{fig:10}
\end{figure}

The $\chi^2$ significance test results for the angular power spectrum are presented in figure~\ref{fig:10}, with the histogram representing the null distribution and the solid vertical line representing the measured $\chi^2$ for the observed BBHs. The data is consistent with the null hypothesis with a $p$-value of 0.061, calculated by fitting a $\chi^2$ function to the null distribution. Note that the relatively small $p$-value is most likely caused by the large ($~3\sigma$) fluctuation at $\ell \roughly 80$. As evident from figure~\ref{fig:9}, this fluctuation does not indicate a clustering signal. Hence, we conclude that \textit{the angular power spectrum does not capture a statistically significant clustering signal in the presently available BBH data}.

\begin{figure}
    \centering
    \includegraphics[width=\columnwidth]{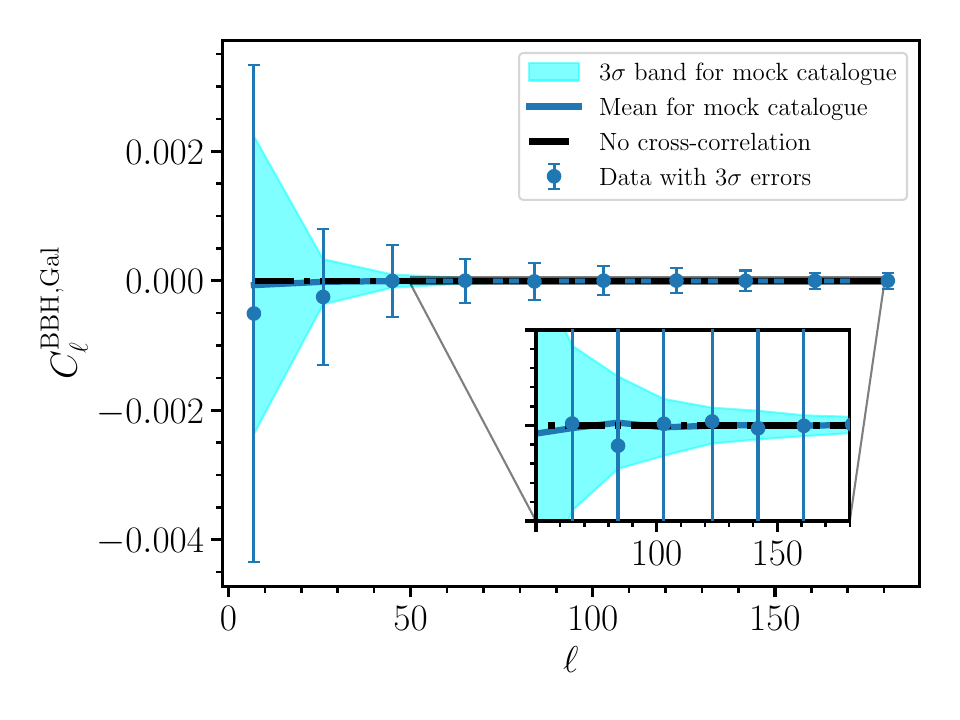}
    \caption{Results of BBH-Galaxy cross-clustering analysis conducted using the cross angular power spectrum as the summary statistic, with a zoomed-in view provided in the inset for better visibility. The plotting scheme is identical to figure~\ref{fig:9}, except the dash-dot line, $y = 0$, represents the cross power spectrum of two uncorrelated data sets. With the present number of BBH detections, no visual evidence exists for a clustering signal in the cross angular power spectrum.}
    \label{fig:11}
\end{figure}

\begin{figure}
    \centering
    \includegraphics[width=\columnwidth]{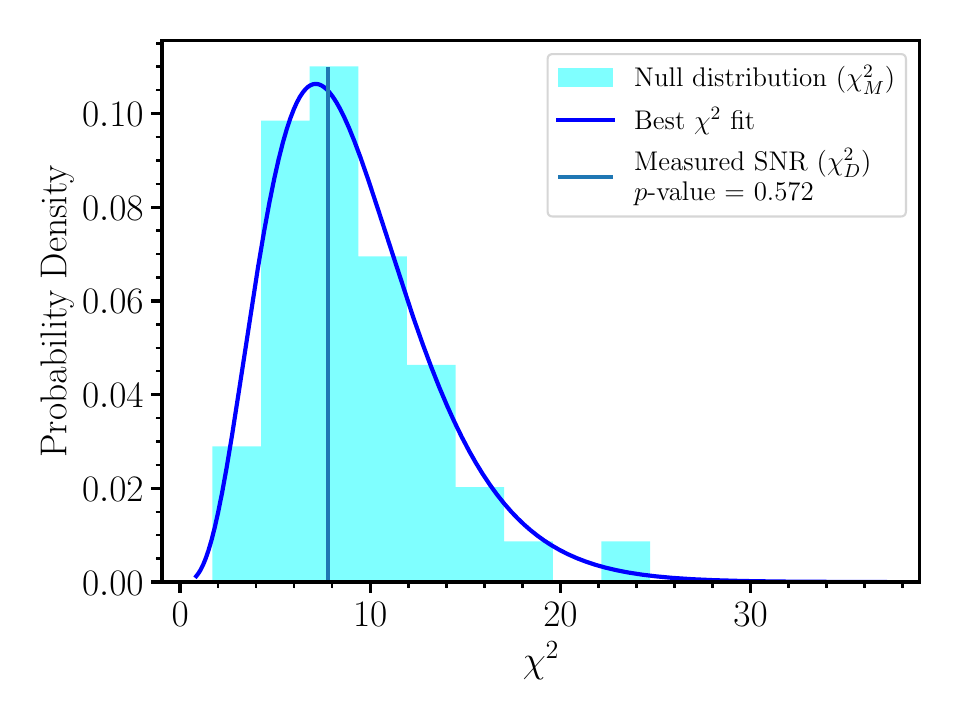}
    \caption{Results of the statistical significance test for the BBH-Galaxy cross angular power spectrum. The plotting scheme is the same as figure~\ref{fig:10}. The data is consistent with the null hypothesis at a $p$-value of 0.572, calculated using the CDF of the best $\chi^2$ fit. No statistical evidence exists for spatial cross-correlation between the presently detected BBHs and the galaxies and quasars from the WSC catalogue.}
    \label{fig:12}
\end{figure}

Figure~\ref{fig:11} shows the cross angular power spectrum measurements between the BBHs and the WSC sources, with the plotting scheme identical to figure~\ref{fig:9}. Again, we plot a zoomed-in version in the inset for better visibility of the shaded band. The mean power spectrum for the mock catalogue is very close to 0 at all scales, as expected from the null hypothesis, which stipulates that the BBHs and WSC sources are spatially uncorrelated (the cross power spectrum is unaffected by shot noise). Even for cross-correlation, the measurement errors on the data are extremely large and exceed the cosmic variance. The $\chi^2$ significance test for the cross angular power spectrum, summarised in figure~\ref{fig:12}, confirms the visual impression given by figure~\ref{fig:11}: the data is consistent with the null hypothesis at a $p$-value of 0.572, implying that \textit{the cross angular power spectrum does not capture statistically significant evidence for spatial cross-correlation between the presently observed BBHs and the WSC sources.}

\subsection{Nearest-neighbour measurements}
\label{subsec:resknn}

\begin{figure*}
    \centering
    \includegraphics[width=\textwidth]{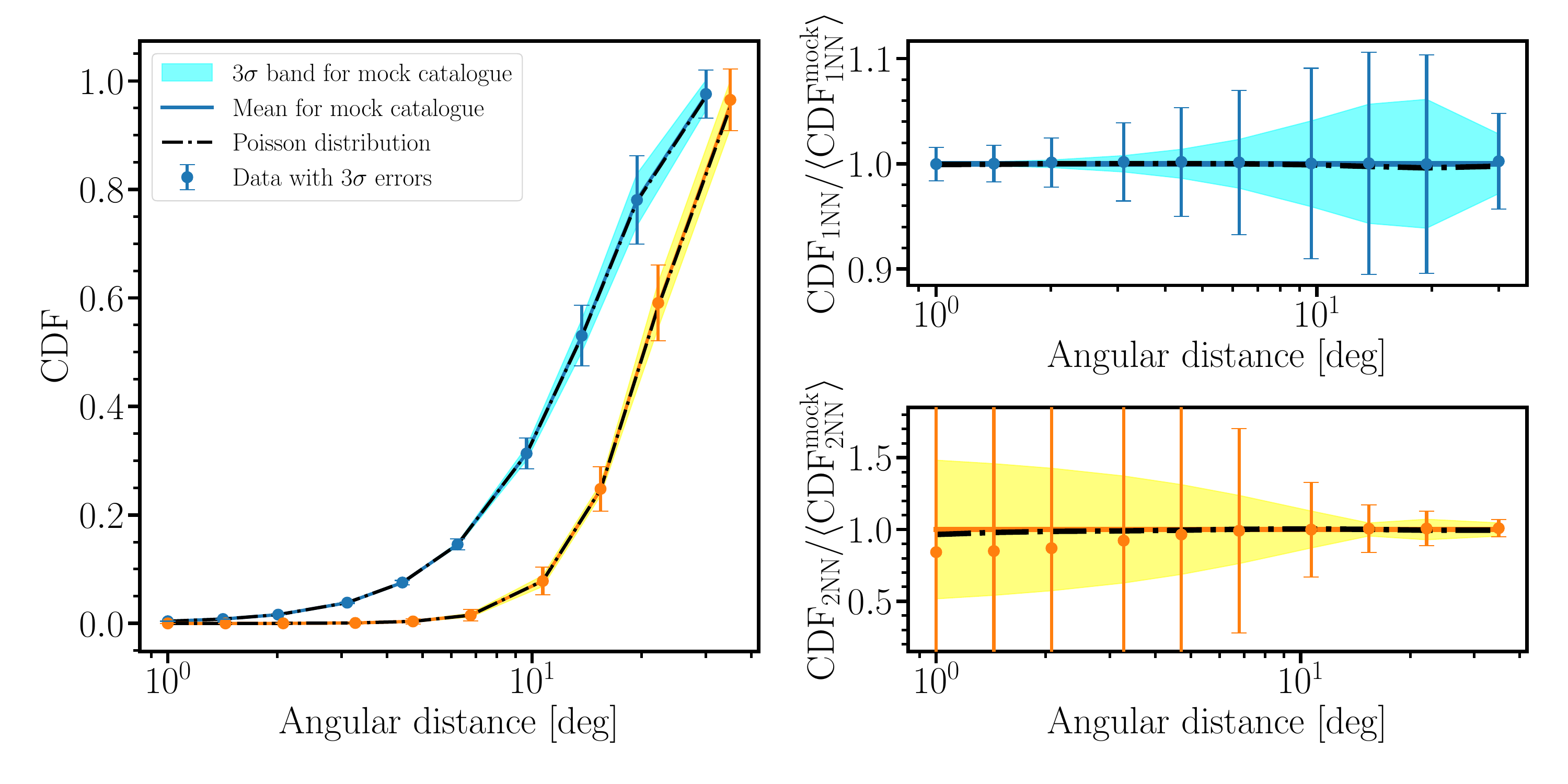}
    \caption{\textit{Left panel}: Results of auto-clustering analysis conducted using the first and second nearest-neighbour cumulative distribution functions as the summary statistics. The plotting scheme is similar to figure~\ref{fig:9}, except the dash-dot line represents the analytic expectation for the $k$NN-CDFs of an unclustered, Poisson-distributed dataset in the sky. Different colours represent different values of the nearest-neighbour index $k$, with blue and orange standing for $k = 1$ and $2$, respectively. \textit{Right panel}: CDFs divided by the mean over the mock catalogue to reduce the dynamic range of the plot. The error bars for the bottom plot have not been shown to the full extent to make the rest of the plot clearer. With the present number of BBH detections, there is no visual evidence for a clustering signal in the $k$NN-CDFs.}
    \label{fig:13}
\end{figure*}

Figure~\ref{fig:13} shows the first two $k$NN-CDFs of the BBHs in blue and orange standing for the first and second neighbours, respectively. The plotting scheme is similar to figure~\ref{fig:9}, except the dash-dot line represents the expectation for the $k$NN-CDFs of an unclustered, Poisson-distributed dataset in the sky. The analytic expression for the $k$NN-CDFs of Poisson distributed points is only a function of $\bar{n}_{\rm BBH} A$, as can be seen from equation~\ref{eq:12} (see also \cite{banerjee_nearest_2021}).

The left panel shows that $\rm CDF_{\rm 1NN}$ is smaller than $\rm CDF_{\rm 2NN}$ at all scales. This is intuitive since, for each query point, the distance to the $k^{\rm th}$ nearest-neighbour is always smaller than the distance to the $(k+1)^{\rm th}$ nearest-neighbour. As a result, for a given scale $\theta$, the fraction of query points with the first nearest neighbour at a distance less than $\theta$ would be larger than those with the second nearest neighbour. This generally holds, i.e., the $k$NN-CDFs always shift towards larger angular scales with increasing $k$.

Since the CDFs span a large range relative to the error bars, it is difficult to visualise the results from the left panel. We plot the CDFs normalised by the mean of the mock catalogue in the right panel of figure~\ref{fig:13}. It is clear from these plots that the mean of the mock catalogue is consistent with the expected value for Poisson distributed data. As a consequence of the uncertainties in the sky localisation of the BBHs, the measurement errors on the CDFs of the observed BBHs are large, even exceeding the variance across the mock catalogue, similar to what was observed for the angular power spectrum in figure~\ref{fig:9}. 

Note how the variance across the mock realisations for the $\rm{CDF}_{\rm 1NN}$ becomes vanishingly small as we approach the smallest angular scales. This happens because at spatial scales much smaller than the mean tracer-tracer separation, the $\rm{CDF}_{\rm 1NN}$ approaches the expected value for a Poisson distribution regardless of the positions of the tracers. Mathematically, as the area $A$ goes to 0, the leading order term in the summation inside the exponential in equation~\ref{eq:12} dominates. Hence, any deviations from the Poisson expression, either due to a clustering signal or due to sampling noise, are exponentially suppressed \cite[see also][]{banerjee_nearest_2021}.

\begin{figure}
    \centering
    \includegraphics[width=\columnwidth]{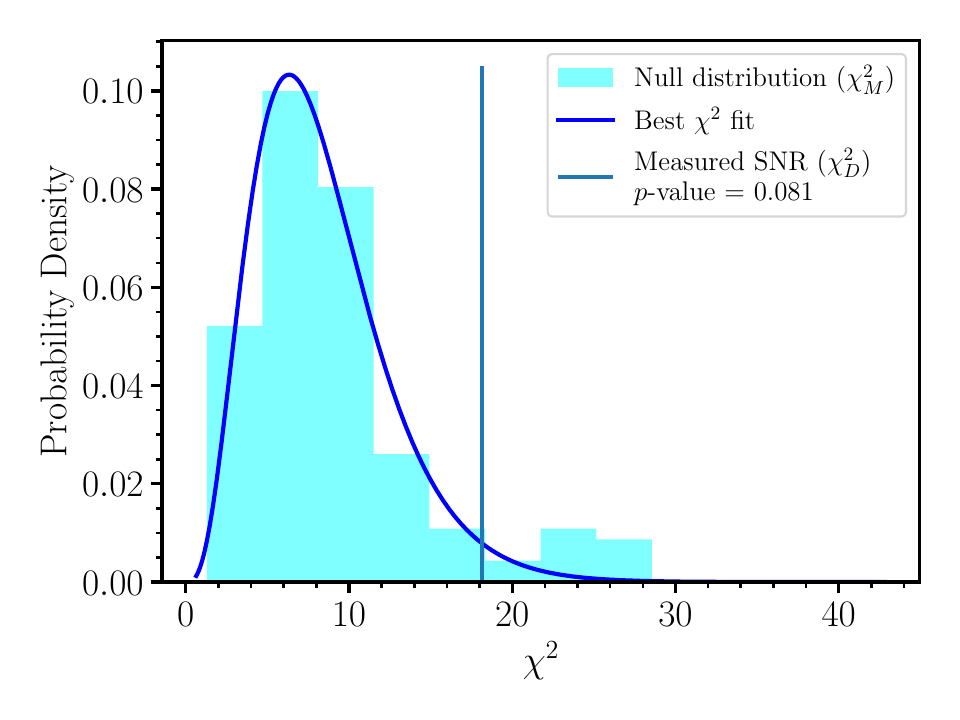}
    \caption{Results of the statistical significance test for the combined first and second nearest-neighbour cumulative distribution functions. The plotting scheme is the same as in figure~\ref{fig:10}. The data is consistent with the null hypothesis at a $p$-value of 0.081, calculated by counting the number of mock realisations with $\chi^2_{M^i} > \chi^2_{D}$ since the best chi-square fit does not characterise the right tail of the distribution well. There is no evidence for a statistically significant clustering signal in the present data.}
    \label{fig:14}
\end{figure}

The data seems to indicate no clustering signal in any of the $k$NN-CDFs. To confirm the visual intuition, we perform a $\chi^2$ significance test using measurements at 5 spatial bins each from the first and second nearest-neighbour distributions and find that the data is consistent with the null hypothesis at a $p$-value of 0.081. Note that in this case, since the null distribution is heavy-tailed and poorly characterised by a $\chi^2$ function, we compute the $p$-value by counting the number of mock realisations with $\chi^2_{M^i} > \chi^2_{D}$. These results are summarised in figure~\ref{fig:14}. We conclude that \textit{the nearest-neighbour distributions do not capture a statistically significant clustering signal in the presently available BBH data}.

\begin{figure*}
    \centering
    \begin{multicols}{2}
        \includegraphics[width=\columnwidth]{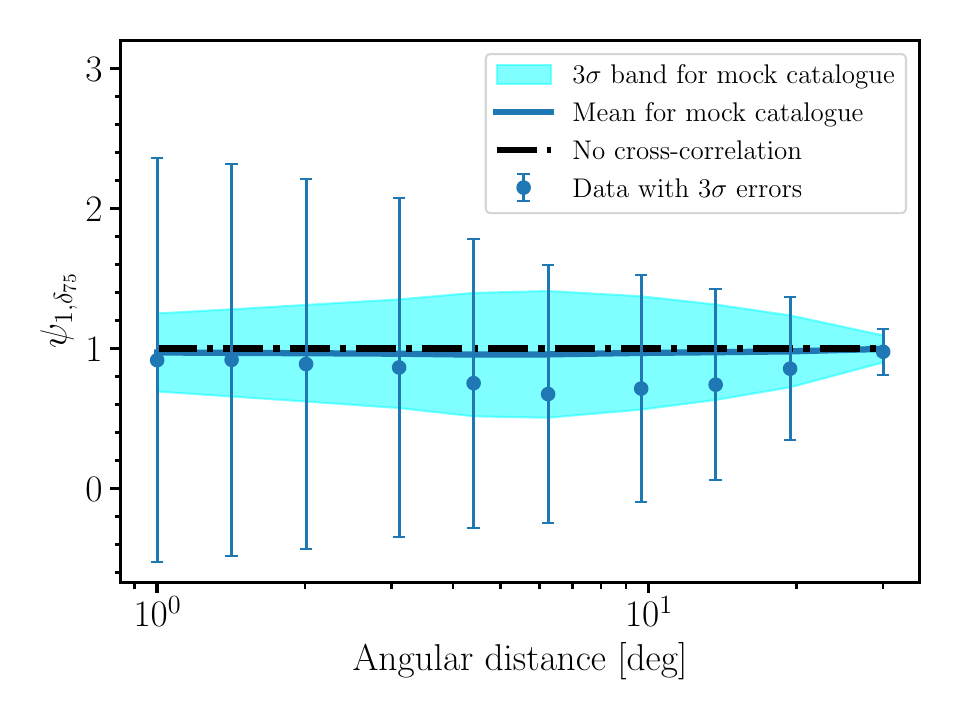}\par 
        \includegraphics[width=\columnwidth]{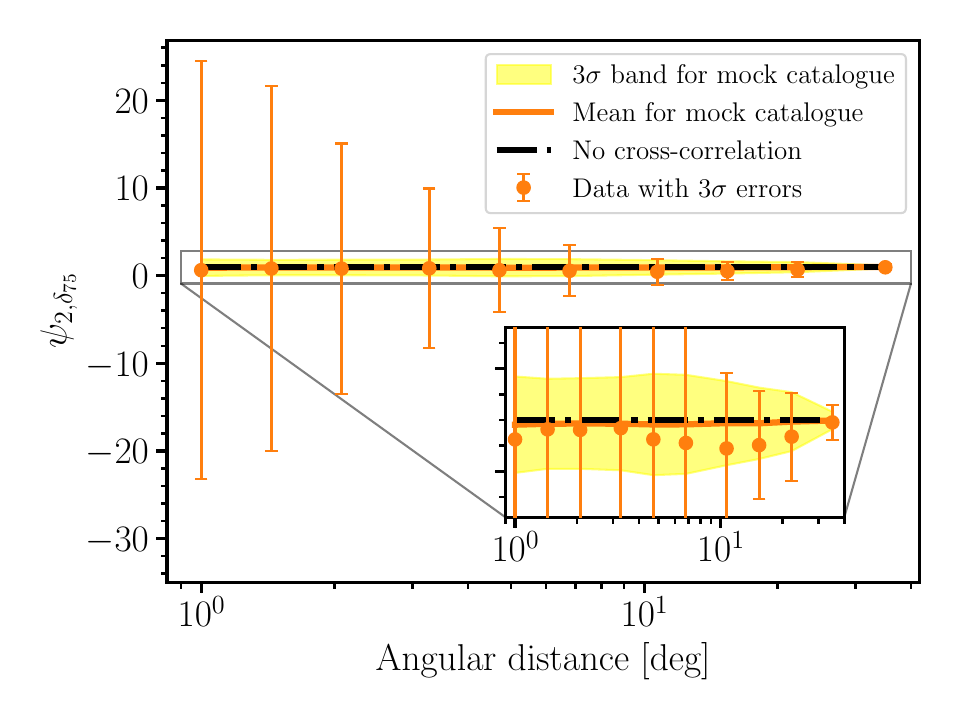}\par 
    \end{multicols}
    \caption{The excess BBH-Galaxy cross-correlation measured by the first (left) and second (right) nearest-neighbour measurements. The plotting scheme is identical to the right panel of figure~\ref{fig:13}, except the dash-dot line here represents the expected excess cross-correlation between two spatially uncorrelated datasets and is identically equal to 1. Since the error bars for the right panel make visualising the rest of the figure difficult, a zoomed-in view is provided in the inset. The plots visually indicate a mild anti-correlation between the BBHs and the WSC catalogue at all angular scales.}
    \label{fig:15}
\end{figure*}

Figure~\ref{fig:15} shows the excess cross-correlation between the BBHs and the quasars and galaxies from the WSC catalogue, as measured by the first and second nearest-neighbour measurements. The plotting and colour schemes are identical to figure~\ref{fig:13}, except the dash-dot line represents the expected excess cross-correlation between two spatially uncorrelated datasets and is identically equal to 1 at all scales. The inset on the right shows a zoomed-in version of the full subplot for better visibility. The mean of the mock catalogue is consistent with 1, as expected from the null hypothesis. As was the case for the other summary statistics, the measurement errors on the data are extremely large even for the excess cross-correlation\footnote{Note that the excess cross-correlation, by definition, cannot take negative values; the error bars extending to negative values is an effect of the choice to plot the mean $\pm$ 3 $\times$ standard deviation, but the actual measurements are always positive.}.  

Interestingly, the nearest-neighbour distributions indicate a mild anti-correlation between the BBHs and the WSC catalogue at all angular scales considered, which is not picked up by the cross angular power spectrum. An anti-correlation between BBHs and large-scale structure would imply that these binary black holes primarily merge in highly isolated, pristine environments such as cosmic voids or in small field galaxies that are not part of larger gravitationally bound groups or clusters. However, even though the anti-correlation seems to manifest in both the nearest-neighbours systematically, extreme care needs to be taken while analysing such plots of the nearest-neighbour distributions; since nearest-neighbour measurements are cumulative, a noise-driven fluctuation on one spatial scale can affect the measurement at nearby scales, and our visual intuition can not be trusted. We need to take this into account by calculating the full covariance matrix before reaching any conclusions. Moreover, the deviation from $\psi = 1$ in each case is well within the limits of cosmic variance as characterised by the mock catalogue. This is likely an effect of sample variance due to the small number of observed BBHs, and we will return to this point in section~\ref{subsec:disc}.

\begin{figure}
    \centering
    \includegraphics[width=\columnwidth]{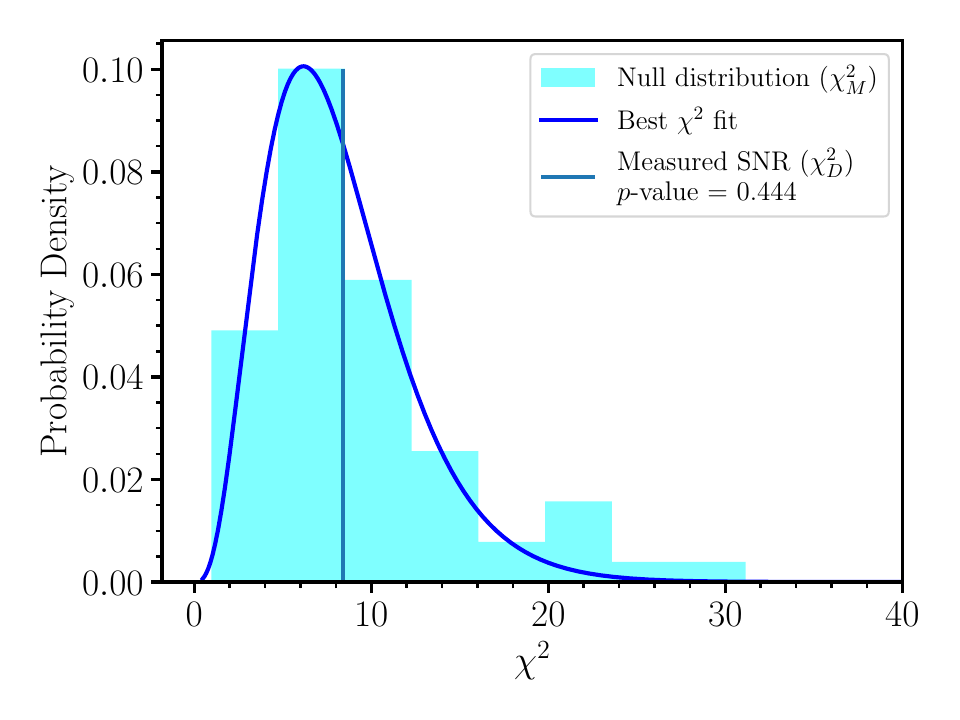}
    \caption{Results of the statistical significance test for the excess BBH-Galaxy cross-correlation as measured using a combination of the first and second nearest-neighbour measurements. The plotting scheme is the same as figure~\ref{fig:10}. The data is consistent with the null hypothesis at a $p$-value of 0.444, calculated by counting the number of mock realisations with $\chi^2_{M^i} > \chi^2_{D}$ since the best chi-square fit does not characterise the right tail of the distribution well. This implies that despite a visual indication for an anti-correlation, there is no statistical evidence for any spatial cross-correlation between the presently detected BBHs and the galaxies and quasars from the WSC catalogue.}
    \label{fig:16}
\end{figure}

We perform a $\chi^2$ significance test using 5 spatial bins each from the first and second nearest-neighbour distributions and find that the data is consistent with the null hypothesis at a $p$-value of 0.444. Similar to the auto-CDFs, the null distribution here is also heavy-tailed. Hence, we compute the $p$-value by counting the number of mock realisations with $\chi^2_{M^i} > \chi^2_{D}$. Figure~\ref{fig:16} shows the summary plot of this analysis. We conclude that \textit{the nearest-neighbour measurements do not capture statistically significant evidence for spatial cross-correlation between the presently observed BBHs and the WSC sources.}

\subsection{Discussion}
\label{subsec:disc}

As we saw in section~\ref{sec:results}, none of the summary statistics considered in this study were able to capture a statistically significant signal, either for the auto-clustering of BBHs or for spatial cross-correlations of BBHs with the large-scale structure of the universe, in the presently available data. Our results are consistent with previous attempts in the literature \cite[see, for example,][]{cavaglia_two-dimensional_2020, mukherjee_cross-correlating_2022, zheng_angular_2023}. What explains these results? If BBHs reside primarily in galaxies, where most stars in the universe live and die, their locations are expected to be clustered and spatially cross-correlated with the observed fluctuations in large-scale structure surveys. 

We believe two aspects of the data conspire to obscure the clustering signal: first, the observed BBHs constitute a statistically small sample that is susceptible to sample variance; with only $\roughly 50$ data points, it is challenging to tell apart a clustered sample from a Poisson-distributed one. Second, with the current sensitivities of the gravitational wave detectors, there is considerable uncertainty in the sky localisation of the BBHs, which tends to smear out any clustering signal at small scales that the nearest-neighbour distributions are the most sensitive to. Of course, there is always the possibility that the null hypothesis is true, in which case, we would not see a clustering signal even if we had perfect observations and a statistically large sample of BBHs. 

The detection of a clustering signal in a small sample such as the one selected for this study would imply that binary black holes reside in extremely biased environments, such as highly dense nodes of the cosmic web or huge cosmic voids. As discussed in section~\ref{subsec:example}, the nearest-neighbour distributions are statistically powerful enough to detect the clustering of rare and highly biased tracers at nonlinear scales. Unfortunately, the non-detection of a clustering signal in the current BBH data does not rule out the possibility of BBHs being highly biased tracers because the uncertainty in the sky localisations would completely wash it out even if a signal existed. Repeating this analysis with better-localised events from future gravitational wave observations would be worthwhile.

\begin{figure}
    \centering
    \includegraphics[width=\columnwidth]{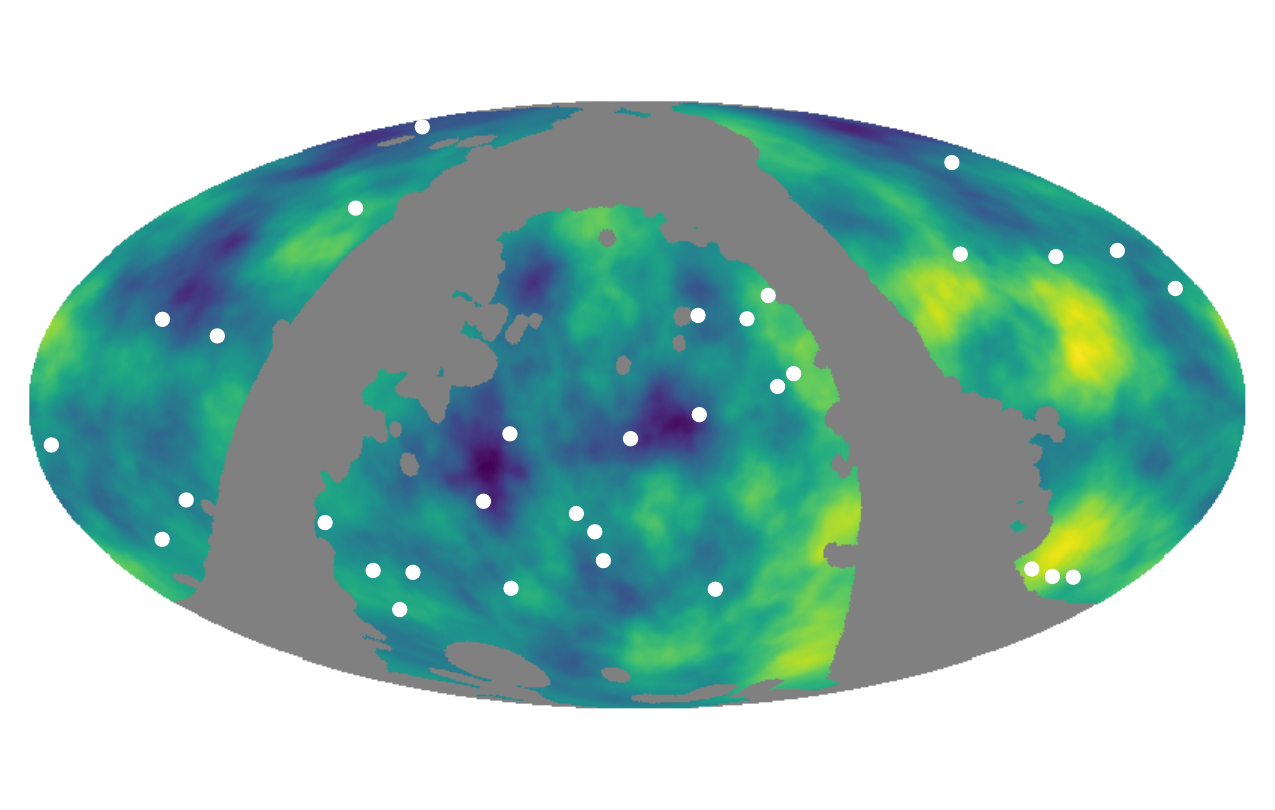}
    \caption{Mollweide projection of the overdensity field for the WSC catalogue smoothed on a $10^{\circ}$ scale using a top-hat filter, with the superimposed white dots representing the most probable positions of the observed BBHs. Warmer colours represent a higher density of galaxies and quasars, while cooler colours represent underdensities. Many of the observed points lie near large-scale underdensities, leading to a slight anti-correlation in the cross-clustering measurements. We believe that this may be due to sample variance since the anti-correlation is not statistically significant, and a significant number of the mock realisations show similar behaviour.}
    \label{fig:17}
\end{figure}

We briefly discussed in section~\ref{sec:results} that the cross-correlation measurements indicate a mild anti-correlation between the observed BBHs and the WSC sources, albeit statistically insignificant. To investigate this further, we plot, in figure~\ref{fig:17}, the most probable sky locations of the observed BBHs on top of the galaxy overdensity field smoothed on $10^{\circ}$ scale using a top-hat filter. Many of the observed BBHs appear to lie near large-scale underdense regions. Due to the small sample size, this is picked up as a mild anti-correlation in the nearest-neighbour distributions. However, this is not evidence for anti-correlation, as established earlier using the $\chi^2$ test. We have further checked that a significant number of the mock realisations show similar behaviour, which is most likely a result of sample variance.

\section{Conclusions and Future Directions}
\label{sec:conc}

In this paper, we developed a framework for quantifying the spatial clustering of sources of gravitational waves and their cross-correlation with the large-scale structure of the universe, using the angular power spectrum and nearest-neighbour distributions as summary statistics. We extended the $k$-nearest-neighbour formalism, originally developed in \cite{banerjee_nearest_2021} and \cite{banerjee_tracer-field_2023} for 3D clustering, to angular clustering in the sky. Our framework implements robust strategies to deal with the extended sky localisation of sources and selection biases associated with gravitational wave detections. It can handle observational systematics due to the presence of masked regions in the sky with unreliable electromagnetic observations. 

We illustrated the statistical power of the nearest-neighbour distributions as measures of spatial clustering of sparsely sampled and highly biased tracers by cross-correlating the overdensity field of the WISE$\times$SuperCOSMOS (WSC) all-sky catalogue with $36$ tracers residing in the highest density regions in the sky. Even with such a small sample size, the first nearest-neighbour distribution captured a statically significant signal at small scales where the angular power spectrum did not. This example demonstrated that the nearest-neighbour distributions can access information in the higher-order correlation functions at small scales where cosmological fluctuations are non-Gaussian.

As a first application to data, we measured the angular power spectrum and nearest-neighbour distributions of the Binary Black Hole (BBH) mergers detected in the first three observation runs of LIGO-Virgo-KAGRA and cross-correlated these sources with galaxies and quasars from the WSC catalogue. We adopted a hypothesis-testing approach to determine the significance of the clustering signal, with the null hypothesis stipulating that BBHs are distributed uniformly in the sky. To mitigate observational biases in the BBH data, we created a catalogue of mock BBHs that statistically reproduce the observed properties of the detected BBHs but are spatially unclustered and uncorrelated with the large-scale structure of the universe. This sample served as a natural control set to compare with the data while testing the null hypothesis.

Using chi-squared distributions to measure statistical deviations from the null hypothesis, we found no evidence for spatial clustering of BBHs or their cross-correlation with large-scale structure in the presently available data. These results are consistent with similar studies in the literature \citep{cavaglia_two-dimensional_2020, mukherjee_cross-correlating_2022, zheng_angular_2023}. We discussed that an absence of a clustering signal is not unexpected, given the small sample size and considerable uncertainty in the sky localisation of the BBHs.

A detection of clustering with so few events would indicate that BBHs reside in extremely biased environments in the universe, such as cosmic web nodes and massive voids. However, a non-detection of this cross-correlation in currently available data does not rule out this scenario since the sky localisation uncertainty smears out the clustering signal at small scales where the measurements are most sensitive. We demonstrated that with well-localised BBHs, the $k$NN tracer-field formalism has the exciting potential to test the possibility of BBHs being highly biased tracers of large-scale structures.

Our framework provides a powerful means to study spatial cross-correlations between continuous fields and rare transient events with uncertain sky localisation in the presence of selection effects and observational systematics. We focused on binary black holes in this work. However, our methods can also be applied to study binary neutron star mergers or neutron star black hole mergers with minor modifications. In addition to gravitational wave sources, other astrophysical transients, such as gamma-ray bursts, are often poorly localised (see \cite{refId1} and references therein). The methods presented in this paper will be useful for conducting multi-messenger studies with these objects. Similarly, the tracer-field correlation formalism discussed here can be applied to conduct cross-correlation studies between large-scale structure and cosmological fields, such as the cosmic microwave background and the cosmological 21 cm neutral hydrogen signal.

While we could not measure the clustering signal in the presently available data on binary black hole mergers, with a statistically significant population of better-localised merger events expected to be detected in future observing runs of LIGO-Virgo-KAGRA and the third generation of gravitational wave detectors \citep{Hall_2019, iacovelli_forecasting_2022, borhanian_listening_2022}, we would have access to even smaller scales where the nearest-neighbour distributions are expected to offer significant gains over a two-point analysis \citep{banerjee_nearest_2021, banerjee_cosmological_2021, banerjee_tracer-field_2023}. Moreover, the larger sample size will also allow us to take a tomographic approach to compute cross-correlations by dividing the data in redshift bins\footnote{Such an approach has been shown to increase the diagnostic power of clustering statistics \cite[see][for a detailed discussion]{calore_cross-correlating_2020}.}. Hence, the techniques developed in this paper would be crucial for measuring the clustering of gravitational wave sources that will be detected in the coming decades. 

In work under preparation, we are conducting forecast studies analysing the angular clustering of BBHs expected to be detected in future observing runs of LIGO and the third-generation detectors, as well as their cross-correlations with forecast galaxy data for stage-IV large-scale surveys. Our preliminary findings suggest that a future detector network including LIGO India \citep{2022CQGra..39b5004S} will detect $\roughly 1.6\times10^4$ BBHs with $1\sigma$ sky localisation area less than $50$ sq. deg. in 10 years of observation, while the third-generation detectors are expected to detect an order of magnitude more number of BBHs with even better sky localisation. We will present the results of a clustering analysis on this forecast data in an upcoming paper (Gupta \& Banerjee 2024 in prep.).

\section*{Acknowledgements}
\label{sec:ack}

The authors would like to thank Aditya Vijaykumar and Susmita Adhikari for helpful comments on the manuscript. We thank Aditya Vijakumar and Prayush Kumar for useful discussions and for their help in creating the mock BBH catalogue. The support and the resources provided by PARAM Brahma Facility under the National Supercomputing Mission, Government of India at the Indian Institute of Science Education and Research; Pune are gratefully acknowledged. The \texttt{iPython} notebook environment \texttt{JupyterLab} \footnote{\url{https://jupyterlab.readthedocs.io/en/latest/index.html}} and the \texttt{python} libraries \texttt{healpy}\footnotemark[12], \texttt{NumPy}\footnote{\url{https://numpy.org/}}, \texttt{pandas}\footnote{\url{https://pandas.pydata.org/}}, \texttt{scikit-learn}\footnotemark[16] and \texttt{SciPy}\footnote{\url{https://scipy.org/}} were used extensively in this work. All plots in this paper were made using \texttt{Matplotlib}\footnote{\url{https://matplotlib.org}}. 

%%%%%%%%%%%%%%%%%%%%%%%%%%%%%%%%%%%%%%%%%%%%%%%%%%
\section*{Data Availability}

All data used in this paper are publicly available; the gravitational wave parameter estimation data and skymaps for the parent BBH catalogue can be found at \url{https://zenodo.org/records/5546663} and the WSC SVM catalogue and mask can be found at \url{http://ssa.roe.ac.uk/WISExSCOS.html}. The data generated in this study, including the mock BBH catalogues, are available upon reasonable request.



\bibliographystyle{mnras}
\bibliography{references} % if your bibtex file is called example.bib

%%%%%%%%%%%%%%%%%%%%%%%%%%%%%%%%%%%%%%%%%%%%%%%%%%

%%%%%%%%%%%%%%%%% APPENDICES %%%%%%%%%%%%%%%%%%%%%

\appendix

\section{BBH properties}
\label{sec:A}
\begin{figure*}
    \centering
    \begin{multicols}{3}
        \subcaptionbox{Component mass\label{fig:A1a}}{\includegraphics[width=1.15\linewidth]{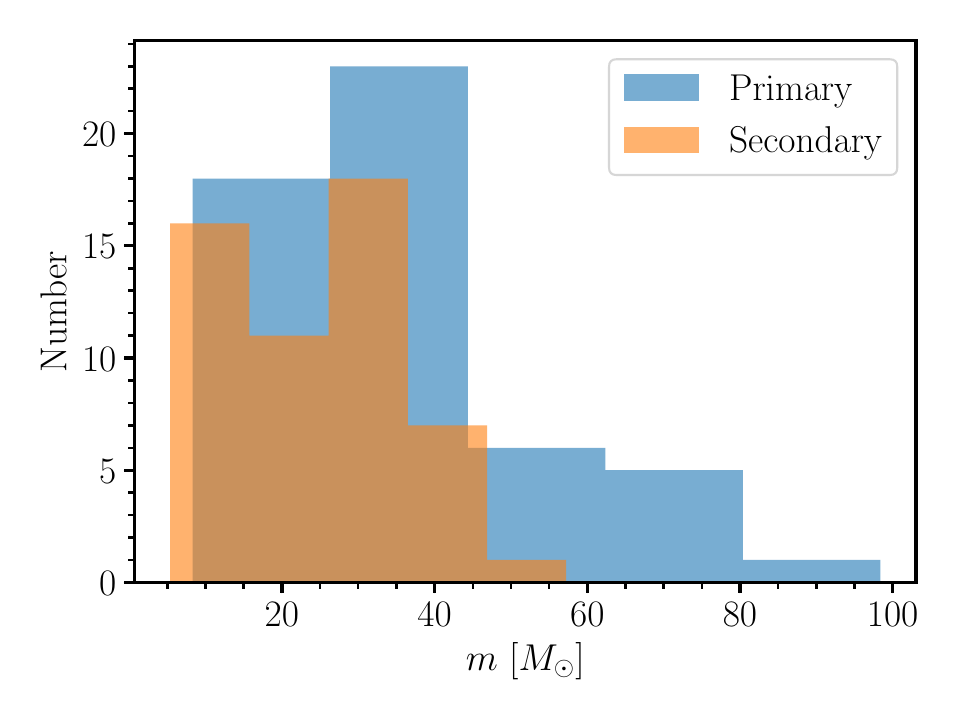}}\par 
        \subcaptionbox{Chirp mass\label{fig:A1b}}{\includegraphics[width=1.15\linewidth]{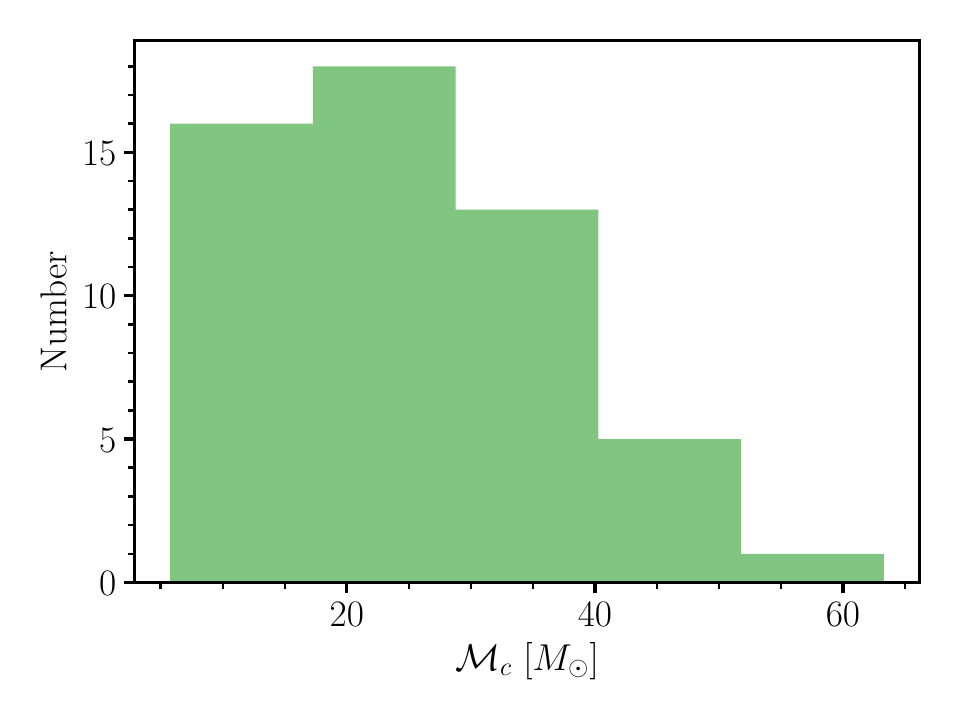}}\par 
        \subcaptionbox{SNR\label{fig:A1c}}{\includegraphics[width=1.15\linewidth]{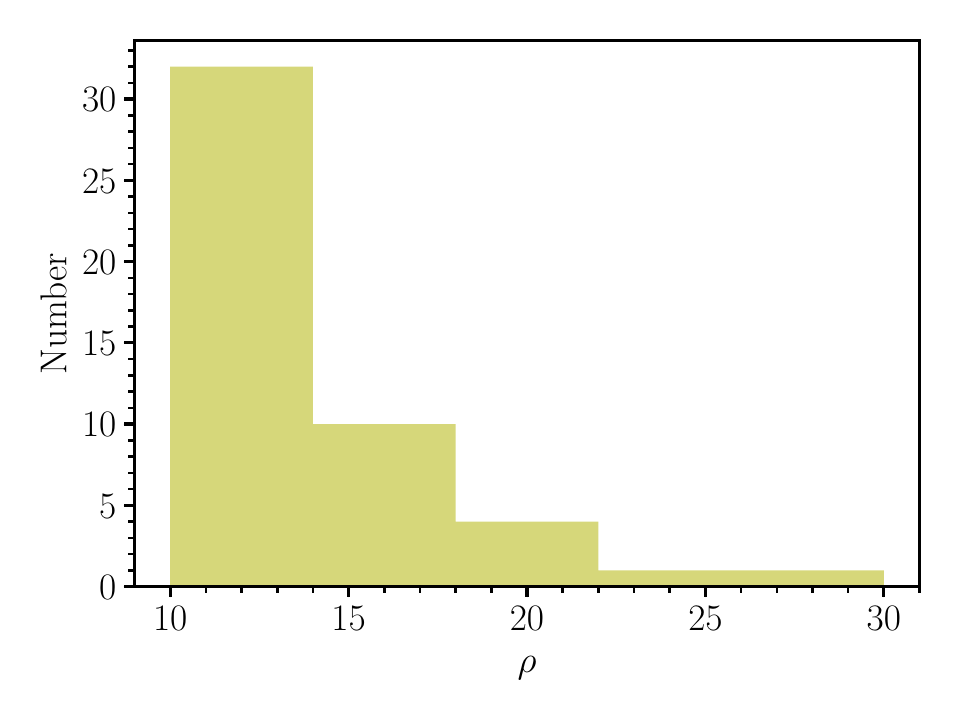}}\par 
    \end{multicols}
    \caption{Distribution of the component masses (\textit{left}), chirp masses (\textit{middle}) and SNRs (\textit{right}) of the observed BBH catalogue. In the left panel, the primary (heavier) BBH mass distribution is shown in blue while the secondary (lighter) BBH mass distribution is shown in orange. The characteristic peak at $\roughly 30 M_{\odot}$ is clearly visible.}
    \label{fig:A1}
\end{figure*}

Figure~\ref{fig:A1} shows the distribution of component masses, chirp masses and network matched-filtered SNRs of the observed BBH catalogue.

\section{Observed vs. Mock BBHs}
\label{sec:C}

\begin{figure*}
    \centering
    \begin{multicols}{3}
        \subcaptionbox{Primary mass\label{fig:C1a}}{\includegraphics[width=1.15\linewidth]{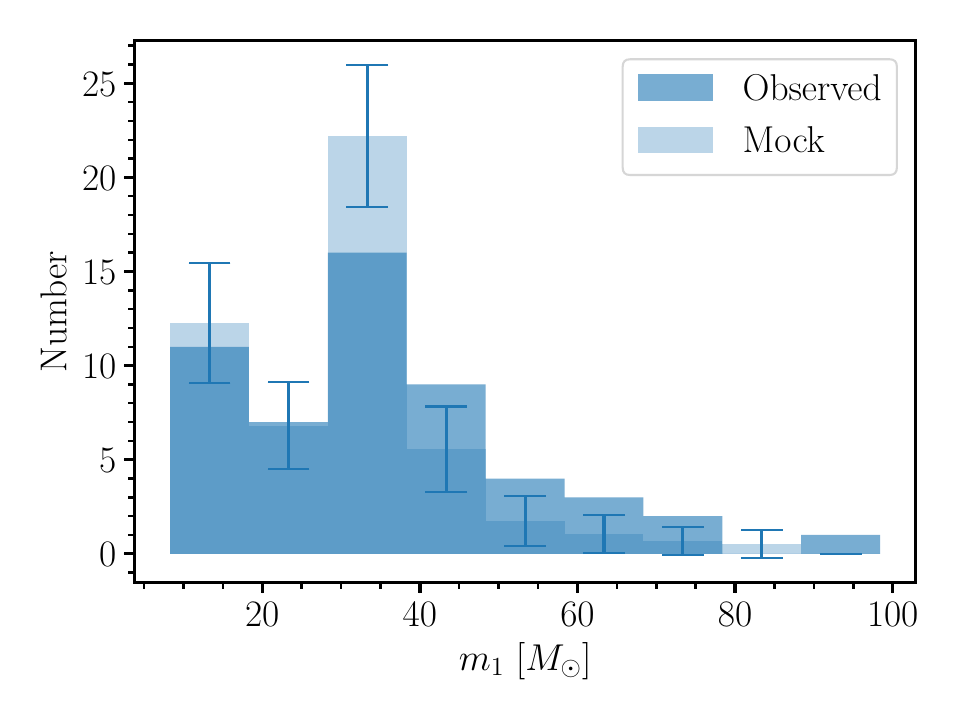}}\par 
        \subcaptionbox{Chirp mass\label{fig:C1b}}{\includegraphics[width=1.15\linewidth]{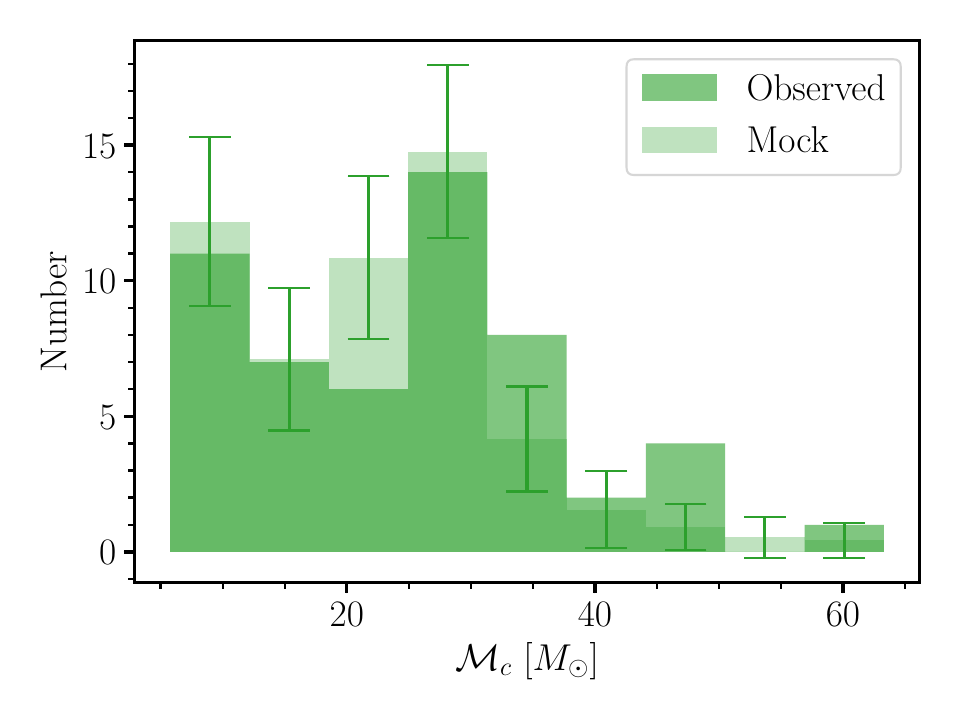}}\par 
        \subcaptionbox{SNR\label{fig:C1c}}{\includegraphics[width=1.15\linewidth]{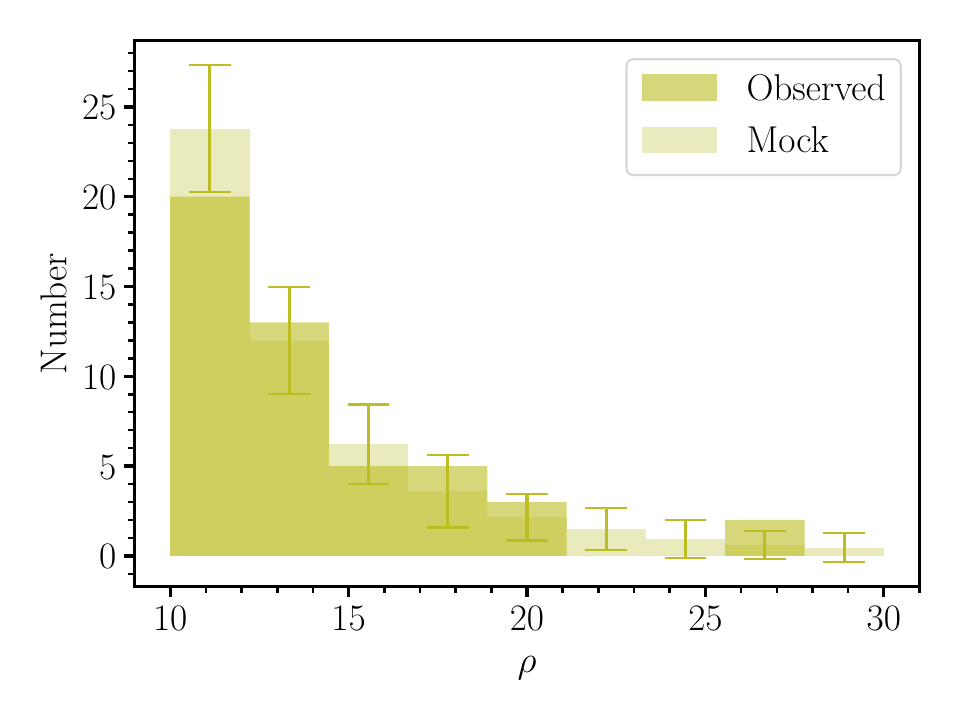}}\par 
    \end{multicols}
    \caption{A comparison of the distribution of the primary masses (\textit{left}), chirp masses (\textit{middle}) and SNRs (\textit{right}) of the observed and mock BBH catalogue. The plotting scheme is the same as figure~\ref{fig:7}. Within error bars, the mock catalogue statistically reproduces the observed data.}
    \label{fig:C1}
\end{figure*}

Figure~\ref{fig:C1} compares the distributions of the primary masses, chirp masses and network matched-filtered SNRs of the observed and mock BBH catalogue. Within error bars, the mock catalogue statistically reproduces the observed data.

\section{Population Models}
\label{sec:B}

For creating the mock BBH catalogue, we assume the \texttt{Power Law + Peak} model for the mass of the primary (heavier) BBH, a power law distribution for the ratio of component masses, and a power law distribution for redshift evolution of merger rate per unit comoving volume per unit source-frame time, as specified in \cite{ligo_scientific_collaboration_population_2023}.

Let $m_1$ denote the mass of the primary black hole, and $q$ denote the mass ratio, such that the mass of the secondary black hole is $qm_1$. The population distribution model for $m_1$ is given by
\begin{align}
    \centering
    \pi(m_1|&\lambda_{\rm peak}, \alpha, m_{\rm min}, \delta_{m}, m_{\rm max}, \mu_{m}, \sigma_{m}) = \notag \\
    &\bigg[(1-\lambda_{\rm peak})\mathcal{P}(m_1|-\alpha, m_{\rm max})+\lambda_{\rm peak}G(m_1|\mu_m, \sigma_m)\bigg] \notag \\
    &\times S(m_1|m_{\rm min}, \delta_{m}) \label{eq:B1}
\end{align}
where $\mathcal{P}(m_1|-\alpha, m_{\rm max})$ is a power law distribution with spectral index $-\alpha$ and high-mass cutoff $m_{\rm max}$, $G(m_1|\mu_m, \sigma_m)$ is a Gaussian distribution with mean $\mu_{m}$ and standard deviation $\sigma_{m}$, and $\lambda_{\rm peak}$ is the mixing fraction that determines the relative importance of the power law and Gaussian components. The lower mass end of the distribution is tapered using a smoothing function $S(m_1|m_{\rm min}, \delta_{m})$ which rises from 0 to 1 over the interval $(m_{\rm min}, m_{\rm min}+\delta_{m})$, given by
\begin{align}
    &S(m_1|m_{\rm min}, \delta_{m}) \notag \\
    &= \begin{cases}
        0 & m<m_{\rm min} \\
        \left[f(m-m_{\rm min}, \delta_{m})+1\right]^{-1} & m_{\rm min} \leq m<m_{\rm min}+\delta_{m} \\
        1 & m \geq m_{\rm min}+\delta_{m}
    \end{cases} \label{eq:B2}
\end{align}
with
\begin{equation}
    f(m, \delta_{m}) = \exp\left(\frac{\delta_{m}}{m} + \frac{\delta_{m}}{m - \delta_{m}}\right) \label{eq:B3}
\end{equation}
The conditional mass ratio distribution is given by a power law, also smoothed at the lower mass end
\begin{equation}
    \pi(q|m_1, \beta_q, m_{\rm min}, \delta_{m}) \propto q^{\beta_{q}}S(qm_1|m_{\rm min}, \delta_{m}) \label{eq:B4}
\end{equation}
The values for the model parameters assumed in this work are taken from the publicly available LVK population analysis results and are summarised in table~\ref{tab:1}.
\begin{table}
    \centering
    \begin{tabular}{|c|c|}
        \hline
        Parameter & Value \\
        \hline\hline
        $\alpha$ & 3.4 \\
        \hline
        $\beta_q$ & 1.08 \\
        \hline
        $m_{\rm min}$ & 5.08 \\
        \hline
        $m_{\rm max}$ & 86.85 \\
        \hline
        $\lambda_{\rm peak}$ & 0.04 \\
        \hline
        $\mu_{m}$ & 33.73 \\
        \hline
        $\sigma_{m}$ & 3.56 \\
        \hline
        $\delta_{m}$ & 4.83 \\
        \hline
    \end{tabular}
    \caption{\texttt{Power Law + Peak} model parameters}
    \label{tab:1}
\end{table}

The power law redshift evolution model parameterises the merger rate density per comoving volume and source-frame time as
\begin{equation}
    \mathcal{R}(z) = \frac{dN}{dV_{\rm c} dt_{\rm s}} = \mathcal{R}_0(1+z)^\kappa \label{eq:B5}
\end{equation}
where $\mathcal{R}_0$ is the merger rate density at $z=0$, $t_{\rm s}$ is the source-frame time, related to observer-frame time as $t_{\rm s} = t_{\rm o}/(1+z)$ due to cosmological redshift. This implies that the observed redshift distribution is
\begin{align}
    \frac{dN}{dz} &= \int dt_{\rm o} \frac{dV_{\rm c}}{dz} \mathcal{R}_0(1+z)^{\kappa-1} \notag \\
    &= t_{\rm obs}\mathcal{R}_0 \frac{dV_c}{dz} (1+z)^{\kappa-1} \label{eq:B6}
\end{align}
where $t_{\rm obs}$ is the total observation time and $\frac{dV_{\rm c}}{dz}$ is the differential comoving volume. The probability distribution function for redshifts is given by normalising equation~\ref{eq:B6}
\begin{equation}
    \pi(z|\kappa, z_{\rm max}) = \frac{\frac{dV_c}{dz} (1+z)^{\kappa-1}}{\int_{0}^{z_{\rm max}} dz \frac{dV_c}{dz} (1+z)^{\kappa-1}} \label{eq:B7}
\end{equation}

where $z_{\rm max}$ is the maximum redshift out to which the population has been created. Note that the constants $\mathcal{R}_0$ and $t_{\rm obs}$ drop out of the expression since they are simply normalisation constants and do not affect the shape of the distribution. However, they do indeed control the total number of mock events to be drawn from the normalised distribution. Although the $z_{\rm max}$ in LIGO analyses is typically taken to be 2.3, in this work, we assume a $z_{\rm max}$ of 0.7 since assuming a lower value of $z_{\rm max}$ is computationally less demanding. We have checked that inputting a higher $z_{\rm max}$ value does not affect our mock catalogue. This is due to the fact that a negligible fraction of injections outside this redshift are detected by our assumed network. We assume $\kappa = 3$ in our analysis.

\section{Smoothing in harmonic space}
\label{sec:D}

Consider a field $\delta(\boldsymbol{\Omega})$ defined in the sky. The expression for the field smoothed on an angular scale $\theta$ is given by
\begin{equation}
    \delta^{\theta}(\boldsymbol{\Omega}) = \frac{1}{2\pi(1-\cos\theta)}\int_{\arccos(\boldsymbol{\hat{\Omega}} \cdot \boldsymbol{\hat{\Omega}}') \leq \theta} d\boldsymbol{\hat{\Omega}}'\delta(\boldsymbol{\hat{\Omega}}') \label{eq:D1}
\end{equation}
The smoothed field is equivalent to the field averaged over spherical caps of angular radius $\theta$. Equation~\ref{eq:D1} can be re-written as
\begin{equation}
    \delta^{\theta}(\boldsymbol{\Omega}) = \int_{\text{All sky}} d\boldsymbol{\hat{\Omega}}'\delta(\boldsymbol{\hat{\Omega}}')W^{\theta}(\boldsymbol{\Omega}', \boldsymbol{\Omega}) \label{eq:D2}
\end{equation}
where $W^{\theta}(\boldsymbol{\Omega}', \boldsymbol{\Omega})$ is the top-hat filter in configuration space, given by
\begin{equation}
    W^{\theta}(\boldsymbol{\Omega}', \boldsymbol{\Omega}) = 
    \begin{cases}
        \frac{1}{2\pi(1-\cos\theta)} & \arccos(\boldsymbol{\hat{\Omega}} \cdot \boldsymbol{\hat{\Omega}}') \leq \theta\\
        0 & \text{otherwise}
    \end{cases} \label{eq:D3}
\end{equation}
The integral in equation~\ref{eq:D3} is computationally expensive to perform in configuration space, but one can use properties of the spherical harmonics and massively speed up the calculation by going to harmonic space. Let the expansions of the smoothed and unsmoothed fields in spherical harmonics be given by
\begin{align}
    \delta(\boldsymbol{\hat{\Omega}}) & = \sum_{\ell m} \alpha_{\ell m} Y_{\ell m}(\boldsymbol{\hat{\Omega}}) \label{eq:D4} \\
    \delta^{\theta}(\boldsymbol{\hat{\Omega}}) & = \sum_{\ell m} \alpha^{\theta}_{\ell m} Y_{\ell m}(\boldsymbol{\hat{\Omega}}) \label{eq:D5}
\end{align}
Since the top-hat filter represents a homogeneous and isotropic smoothing kernel that is only a function of the central angle $\theta$ between $\boldsymbol{\hat{\Omega}}$ and $\boldsymbol{\hat{\Omega}}'$, it can be expanded in terms of the Legendre polynomials $P_{\ell}(\cos\theta)$ as:
\begin{equation}
    W^{\theta} = \sum_{\ell} b_{\ell} P_{\ell}(\cos\theta) \label{eq:D6}
\end{equation}
It can be shown, by substituting equations~\ref{eq:D4} to~\ref{eq:D6} in equation~\ref{eq:D2}, that the spherical harmonic expansion coefficients $\alpha^{\theta}_{\ell m}$ of the smoothed field are given by the product of the $\alpha_{\ell m}$ of the unsmoothed field and the Legedre expansion coefficients $b_{\ell}$ of the top-hat filter \citep{devaraju_understanding_2015}\footnote{Note that we have an additional factor of $4\pi$ not present in the expression derived by \cite{devaraju_understanding_2015}, since we use a normalised definition for the top-hat filter that already incorporates an extra factor of $1/4\pi$ that appears in their equivalent of equation~\ref{eq:D2}.}:
\begin{equation}
    \alpha^{\theta}_{\ell m} = 4\pi\frac{b_{\ell}} {2\ell+1}\alpha_{\ell m} \label{eq:D7}
\end{equation}
This expression makes sense since computing the integral in equation~\ref{eq:D3} is equivalent to performing a \textit{spatial convolution} on the surface of a sphere. A convolution in configuration space corresponds to a product in harmonic space. It is to be noted that for a general inhomogeneous smoothing kernel, however, spatial smoothing is not equivalent to a convolution since each point on the sphere has a different kernel. For such cases, equation~\ref{eq:D7} would be different. See chapter 2 of \cite{devaraju_understanding_2015} for a nice discussion on the topic.

In practice, we compute the $\{\alpha_{\ell m}\}$ for the unsmoothed field using the \texttt{healpy} package. The $\{b_{\ell}\}$ for the top-hat function are computed as follows. By definition,
\begin{align}
    b_{\ell} &= \frac{2\ell+1}{2}\int_{-1}^{1}W^{\theta}P_{\ell}(\cos\theta)d(\cos\theta) \notag \\
    &= \frac{2\ell+1}{2}\frac{1}{2\pi(1-\cos\theta)}\int_{\cos\theta}^{1}P_{\ell}(\cos\theta)d(\cos\theta) \label{eq:D8}
\end{align}
Using the recursion relation
\begin{equation}
    P_{\ell}(x) = \frac{1}{2\ell+1}\frac{d}{dx}\left[P_{\ell+1}(x)-P_{\ell-1}(x)\right] \label{eq:D9}
\end{equation}
we get
\begin{equation}
    b_{\ell} = \frac{1}{4\pi(1-\cos\theta)}\left[P_{\ell-1}(\cos\theta)-P_{\ell+1}(\cos\theta)\right]\label{eq:D10}
\end{equation}

%%%%%%%%%%%%%%%%%%%%%%%%%%%%%%%%%%%%%%%%%%%%%%%%%%

% Don't change these lines
\bsp	% typesetting comment
\label{lastpage}
\end{document}